\documentclass[aps,prd,superscriptaddress,onecolumn, nofootinbib]{revtex4-2}

\usepackage{blindtext}
\usepackage{cancel}
\usepackage{mathtools, mathalfa, amssymb, amsmath, amsfonts, amsthm, bm}
\usepackage{graphicx}
\usepackage{tikz}
\usepackage{multirow}
\usepackage[table, dvipsnames]{xcolor}
\usepackage{color}
\usepackage{framed}
\usepackage{aas_macros}
\usepackage{amsthm}

\usepackage{appendix}
\usepackage{float}
\usepackage{subcaption}
\usepackage{xcolor}
\usepackage{orcidlink}
\usepackage{algorithm}
\usepackage{algpseudocode}
\usepackage{circledsteps}
\pgfkeys{/csteps/fill color=yellow}

\usepackage{hyperref}
\hypersetup{
    colorlinks = true,
    citecolor  = blue,
    linkcolor = blue,
    urlcolor  = blue,
    linkbordercolor = white
}

\graphicspath{{./figures/}}
\usepackage[bottom]{footmisc}

\usepackage{cleveref}

\newcommand{\dd}{\mathrm{d}}
\newcommand{\ii}{\mathrm{i}}

\begin{document}
    \title{Asymptotic evaluation of multidimensional oscillatory integrals}
    \author{Samuel Crew
    \orcidlink{0000-0002-6136-0910}} 
    \email{samuel.crew@qperfect.io}
    \affiliation{QPerfect, 23 Rue du Loess, 67200 Strasbourg, France}

    \author{Job Feldbrugge
    \orcidlink{0000-0003-2414-8707}}
    \email{jobfeldbrugge@ntu.edu.tw}
    \affiliation{Leung Center for Cosmology and Particle Astrophysics, National Taiwan University, No. 1, Sec. 4, Roosevelt Road, Taipei 10617, Taiwan}
    \affiliation{Institute for Advanced Study, 1 Einstein Drive, Princeton, NJ 08540, USA}

    \begin{abstract}
        Oscillatory integrals appear throughout physics, from diffraction and gravitational lensing to real-time path integrals. Their asymptotic evaluation is complicated by the need to determine which real and complex saddle points contribute, and how they are connected by Stokes phenomena. In multiple dimensions, this traditionally requires computing the steepest descent manifolds in a high-dimensional complex space. Here, we develop an algebraic alternative for integrals with polynomial and rational exponents. Using integration by parts and Brieskorn reduction, we construct a finite-dimensional Gauss-Manin system that generates the asymptotic expansions of all saddle points. Its Borel transform yields a Fuchsian system, whose monodromy determines the Stokes constants and hence the adjacency of the saddles. Combined with the recent South-East rule, this identifies the contributing saddles and enables superasymptotic and hyperasymptotic approximations without computing gradient flows. We demonstrate the method for the parabolic umbilic diffraction integral and a Lorentzian lens as examples, obtaining agreement to machine precision. The methods are implemented in an accompanying open source \verb|Julia| package \verb|IntegralResurgence.jl|.
    \end{abstract}
    
    \maketitle

    \section{Introduction}\label{sec:introduction}
    Oscillatory integrals of the form
    \begin{align}
        \Psi = \left(\frac{k}{2\pi \mathrm{i}}\right)^{d/2}\int_{\mathbb{R}^d}e^{\mathrm{i} k f(\bm{x})}\mathrm{d}\bm{x}\,, \label{eq:OrigionalEquation}
    \end{align}
    with integration variable $\bm{x}=(x_1, x_2, \dots, x_d)$, a real exponent $f$ and a parameter $k$, are ubiquitous in theoretical physics. They describe, for example, the diffraction of light near caustics \cite{Nye:1999, Berry:1990b}; the wave optics of gravitational lenses \cite{Schneider:1992, Feldbrugge:2020arXiv200801154F, Feldbrugge:2023b, Feldbrugge:2026}; and the strong-field and attosecond physics of atoms in intense laser fields \cite{salieres:2001, Pisanty:2020, Weber:2025}. In particular, discretised real-time Feynman path integrals in quantum mechanics, quantum field theory and quantum gravity often take this form, with a large number of integration variables \cite{Feynman:1948, Feynman:1965, Feldbrugge:2017, Feldbrugge:2023}. At the same time, these integrals are notoriously hard to evaluate. They converge only conditionally, their value emerges from delicate cancellations of rapidly oscillating contributions, and the computational cost of direct numerical integration grows rapidly with both the parameter $k$ and the dimension $d$. The same oscillations lie at the root of the sign problem in quantum Monte Carlo simulations \cite{Loh:1990, Pan:2022} and the conformal factor problem in quantum cosmology \cite{Gibbons:1978, Hawking:1979, Mazur:1990}, which has motivated the deformation of path integrals onto Lefschetz thimbles \cite{Witten:2010, Cristoforetti:2013, Fujii:2013,Feldbrugge:2017}.

    For large $k$, the integral is classically approximated with the method of stationary phase. The integral localises on the critical points $\bm{x}_j$ of the exponent, and each critical point contributes an exponential $e^{\ii k f(\bm{x}_j)}$ multiplied by an asymptotic series in $1/k$ \cite{JonesKline:1958, Bleistein:1969, Ursell:1980, Fedoryuk:1977, Wong1989, ArnoldVG:1985, ArnoldVG:1988}. These series generically diverge, as do the perturbative expansions of quantum field theory \cite{Dyson:1952}. Far from being a defect, the divergence encodes the global structure of the integral. In the theory of resurgence, pioneered by Dingle \cite{Dingle:1973} and Ecalle \cite{Ecalle:1981} and developed for integrals and differential equations by Berry, Howls and Olde Daalhuis \cite{Berry:1990, Berry:1991, Howls:1992, Howls:1997, Daalhuis:1996, Daalhuis:1998, Daalhuis:1998b}, the late terms of the series around one critical point are determined by the early terms of the series around the critical points it is \textit{adjacent} to, weighted by integer Stokes constants. Truncating each series at its smallest term yields an exponentially accurate \textit{superasymptotic} approximation. Re-expanding the remainders in terms of the series of the adjacent critical points yields \textit{hyperasymptotic} approximations of ever increasing, in principle unlimited, accuracy \cite{Berry:1990, Berry:1991, Bennett:2018}. In recent decades, resurgence has grown into a central tool in theoretical physics, connecting perturbative and non-perturbative effects in quantum mechanics and quantum field theory \cite{Dunne:2012, Basar:2017, Serone:2017, Aniceto:2018bis, Marino:2019wra, crew2024resurgent}.

    Shortly after introducing hyperasymptotics, Berry and Howls anticipated the extension of the theory to multidimensional integrals \cite{Berry:1991}:
    \begin{quote}
        \textit{``There is the extension to integrals with more than one variable. As well as being necessary to beat the halting barrier and approximate the hyperterminants, as just discussed, hyperasymptotics of multiple integrals would be useful in physics. Double integrals occur naturally in diffraction theory, where large $k$ corresponds to the geometrical-optics limit and small exponentials represent complex rays. Infinite-dimensional integrals occur in statistical mechanics and quantum field theory, where small exponentials represent instantons. It is tempting to regard hyperasymptotics as a model for the feeble interaction of particles, represented by saddles. Each lowest saddle-point contribution gives the `bare mass' of the corresponding particle; higher-order terms give the `local renormalization' of the mass; the renormalization diverges, and hyperasymptotics describes the ultimate breakdown of locality, when the particles can no longer be considered in isolation.''}
    \end{quote}
    Howls subsequently developed the hyperasymptotics of multidimensional integrals, reducing the integrals over steepest descent manifolds in $\mathbb{C}^d$ to integrals in a single complex Borel plane \cite{Howls:1997, Delabaere:2002}. Yet, more than three decades later, the vision of Berry and Howls has rarely been realized in practice beyond a single dimension. The reason is that the resurgent approximation of an integral requires three pieces of \textit{global} information, which are hard to obtain in more than one dimension: (i) the asymptotic series of all critical points, (ii) the adjacency of the critical points, encoded in the Stokes constants, and (iii) the relevance of the critical points, \textit{i.e.}, which real and complex critical points actually contribute to the integral over $\mathbb{R}^d$. In one dimension, the latter two follow from a plot of the steepest descent contours. In higher dimensions, they traditionally require the computation of steepest descent and ascent manifolds in $\mathbb{C}^d$ by integrating gradient flows \cite{Kaminsk:i1994, Shoji:2026}, which rapidly becomes intractable as the dimension grows. Consequently, previous applications of hyperasymptotics to multidimensional integrals assumed the relevant critical points and their adjacency to be known a priori \cite{Howls:1997, Delabaere:2002}.

    In this paper, we show that all three ingredients can be obtained \textit{algebraically}, without computing a single flow line. Our starting point is the observation that integration by parts relates the moments of the integrand. The space of integrands modulo total derivatives, the twisted de Rham cohomology of the exponent \cite{Schwarz:2009}, is finite-dimensional, with a dimension equal to the number of critical points $n$. Reducing the integrand with the Brieskorn reduction \cite{Brieskorn:1970} in a good basis \cite{Saito:1989, Saito:2013, Sabbah:1999, Schulze:2003}, the moments satisfy a linear system of $n$ ordinary differential equations in $k$, the Gauss-Manin system. Physicists will recognize this procedure from the reduction of Feynman integrals to master integrals with integration-by-parts identities \cite{Chetyrkin:1981qh, Laporta:2000dsw}, which has recently been formulated in terms of twisted cohomology and intersection theory \cite{Mizera:2017rqa, Mastrolia:2018uzb, Frellesvig:2019uqt, Weinzierl:2022eaz}. The Gauss-Manin system turns the analysis of a $d$-dimensional integral into the analysis of an $n$-dimensional linear differential equation, with an irregular singularity at $k=\infty$:
    \begin{itemize}
        \item The asymptotic series of all critical points follow from a simple recursion relation, to any order and in arbitrary precision.
        \item The Borel transform of the moments satisfies a Fuchsian system with simple poles at the critical values. Its solutions can be continued analytically along any path in the Borel plane, either by transporting the solution numerically or with hyperlogarithms, yielding the Stokes constants as integers with certified precision.
        \item Combined with the critical values, the Stokes constants determine the relevant critical points through the South-East rule \cite{Aniceto:2026SouthEast}, a simple geometric algorithm in the Borel plane that replaces the computation of the intersection numbers of Picard-Lefschetz theory.
        \item With the relevant critical points, their asymptotic series and their adjacency at hand, the integral can be approximated to first order with the superasymptotic expansion or to higher orders using the hyperasymptotic expansion.
    \end{itemize}
    Together, these steps form a pipeline that takes a polynomial or rational exponent as input and returns an approximation of the oscillatory integral to any desired precision. For mathematicians, the construction connects the Brieskorn lattice and the Birkhoff problem for tame polynomials with the resurgence of Écalle and the hyperasymptotics of Berry and Howls. In this picture, the Stokes constants are the intersection numbers of vanishing cycles, which we compute from the monodromy of a Fuchsian system. For physicists, it offers a practical and dimension-independent method to evaluate oscillatory integrals in the regime where direct numerical integration fails, and to identify the complex saddle points, such as complex rays and instantons, that contribute to them.

    Alongside this paper, we release the \verb|Julia| package \verb|IntegralResurgence.jl|, which implements these methods. Building on the \verb|Singular.jl| Julia package for the evaluation of the Gauss-Manin system \cite{Schulze:2003}, the package computes the critical points, the asymptotic series, Borel-Pad\'e approximants and the Stokes constants, visualizes the adjacency graph in the Borel plane, determines the relevant critical points with the South-East rule, and evaluates the superasymptotic approximation, truncating each series at the distance to the nearest adjacent singularity and estimating the remaining error. We hope that the \verb|IntegralResurgence.jl| package will make the mathematics developed in this paper available for practical problems in mathematics, physics and astronomy. The package is available at \url{https://github.com/jfeldbrugge/IntegralResurgence.jl}.
    We illustrate the method with two examples: the parabolic umbilic diffraction integral, a genuinely two-dimensional integral with an exponent that is unbounded from below, and the Lorentzian lens model of wave optics. In both cases, the method identifies the relevant complex saddle points and their adjacency, and the superasymptotic approximation reproduces the integral to machine precision.

    The paper is organized as follows. In \cref{sec:Picard-Lefschetz}, we review the Picard-Lefschetz formula. In \cref{sec:GaussManin}, we derive the Gauss-Manin system with the Brieskorn reduction, and in \cref{sec:asymptotic}, we develop the asymptotic series of the critical points. In \cref{sec:Borel}, we transform the Gauss-Manin system to the Fuchsian system in the Borel plane and evaluate the Stokes constants. In \cref{sec:approximation}, we combine the adjacency relations with the South-East rule, superasymptotics and hyperasymptotics to approximate the integral. We illustrate the method in \cref{sec:Example} and conclude in \cref{sec:Conclusion}. In the appendices, we discuss linear differential equations with an irregular singularity (\cref{ap:ODE}) and give a more formal account of twisted de Rham cohomology and the Brieskorn reduction (\cref{ap:twisted}).

    \section{The Picard-Lefschetz formula}\label{sec:Picard-Lefschetz}
    In this section, we introduce the integrals along the Lefschetz thimbles and set the notation used throughout the paper. We consider the integral \cref{eq:OrigionalEquation} with a polynomial exponent $f \in \mathbb{R}[\bm{x}]$. Rational exponents can be brought into this form by a polynomial lift, as we illustrate by example for the Lorentzian lens in \cref{sec:lens}. We assume that $f$ is a Morse function, \textit{i.e.}, that its critical points
    \begin{align}
        \nabla f(\bm{x}_j) = \bm{0}\,, \quad j = 1, \dots, n\,,
    \end{align}
    are non-degenerate, $\det \mathcal{H}f(\bm{x}_j) \neq 0$ with the Hessian $\mathcal{H}f$, and that their critical values
    \begin{align}
        \lambda_j = f(\bm{x}_j)
    \end{align}
    are distinct. We moreover assume that $f$ is tame, \textit{i.e.}, that it has no critical points at infinity \cite{Sabbah:1999}. The critical points are generally complex, $\bm{x}_j \in \mathbb{C}^d$. As $f$ has real coefficients, they are either real or come in complex conjugate pairs, $(\bm{x}_j, \bar{\bm{x}}_j)$, with complex conjugate critical values. We will refer to the critical points as saddle points, irrespective of their Morse index. Many aspects of the present analysis generalize to complex valued polynomials $f \in \mathbb{C}[\bm{x}]$.

    \paragraph{Lefschetz thimbles.}
    Following Picard, Lefschetz and Pham \cite{Pham1965FormulesDP, pham2011singularities, ArnoldVG:1988}, we analytically continue the integrand to $\mathbb{C}^d$ and deform the integration domain $\mathbb{R}^d$ to a contour along which the integrand no longer oscillates. The exponent $\ii k f(\bm{x})$ is a holomorphic function, whose real part
    \begin{align}
        h(\bm{x}) = \mathrm{Re}[\ii k f(\bm{x})]
    \end{align}
    sets the magnitude of the integrand and whose imaginary part sets its phase. Consider the gradient flow of $h$ with respect to the Euclidean metric on $\mathbb{C}^d \cong \mathbb{R}^{2d}$,
    \begin{align}
        \frac{\dd \bm{x}}{\dd s} = \mp \overline{\frac{\partial\, (\ii k f)}{\partial \bm{x}}}\,,
    \end{align}
    where the minus (plus) sign corresponds to the flow of steepest descent (ascent). By the Cauchy-Riemann equations, the phase $\mathrm{Im}[\ii k f(\bm{x})]$ is conserved along the flow, while the magnitude $h(\bm{x}) = \text{Re }[\ii k f(\bm{x})]$ decreases (increases) monotonically. The Lefschetz thimble $\mathcal{J}_j$ of the saddle point $\bm{x}_j$ is the union of all paths of steepest descent emanating from $\bm{x}_j$, and the steepest ascent manifold $\mathcal{K}_j$ is the union of all paths of steepest ascent. Both are real $d$-dimensional manifolds in $\mathbb{C}^d$, which intersect at the saddle point. Along the thimble $\mathcal{J}_j$, the phase of the integrand is constant, $H[\bm{x}] = \mathrm{Im}[\ii k f(\bm{x})] = \mathrm{Im}[\ii k \lambda_j]$, and its magnitude decays exponentially away from $\bm{x}_j$. The integrals along the thimbles therefore converge absolutely and are free of oscillations. For later use, we define the thimble integrals
    \begin{align}
        \Psi^{(j)}(k) = \left(\frac{k}{2\pi \mathrm{i}}\right)^{d/2} \int_{\mathcal{J}_j} e^{\mathrm{i} k f(\bm{x})}\mathrm{d}\bm{x}\,.\label{eq:thimble}
    \end{align}

    \paragraph{The Picard-Lefschetz formula.}
    The thimbles form a basis of the integration contours along which the integrand decays (see \cref{ap:twisted} for a more formal account). Consequently, the original integration domain can be deformed into a sum of thimbles, yielding the Picard-Lefschetz formula
    \begin{align}
        \Psi = \sum_{j=1}^n n_j\, \Psi^{(j)}\,,\label{eq:PicardLefscehtz}
    \end{align}
    with the intersection numbers
    \begin{align}
        n_j = \langle \mathbb{R}^d, \mathcal{K}_j\rangle \in \mathbb{Z}\,.
    \end{align}
    The intersection number counts, with orientation, the number of times the steepest ascent manifold $\mathcal{K}_j$ intersects the original integration domain. This follows from the duality $\langle \mathcal{J}_i, \mathcal{K}_j\rangle = \delta_{ij}$ of the descent and ascent manifolds, which holds for generic phases of $k$. A saddle point is \textit{relevant} when $n_j \neq 0$, and \textit{irrelevant} when $n_j = 0$. Every real saddle point is relevant: the steepest ascent manifold of a real saddle point intersects the real plane exactly once, at the saddle point itself, such that $n_j = \pm 1$. A complex saddle point is relevant when its steepest ascent manifold reaches back to the real plane. In general, the relevant complex saddle points cannot be identified from their location in $\mathbb{C}^d$ or from the size of their contribution alone. Including an irrelevant saddle point in an approximation often leads to an exponentially large error.

    \paragraph{Orientation and normalization.}
    The intersection numbers depend on the orientation of the thimbles. We orient $\mathcal{J}_j$ such that the leading term of the method of steepest descent reads
    \begin{align}
        \Psi^{(j)}(k) = \frac{e^{\ii k\lambda_j}}{\sqrt{\det \mathcal{H}f(\bm{x}_j)}}\left[1 + \mathcal{O}(k^{-1})\right],\label{eq:normalization}
    \end{align}
    with the principal branch of the square root for complex saddle points. For a real saddle point with Morse index $\nu_j$ (the number of negative eigenvalues of the Hessian), the method of stationary phase yields the contribution $e^{\ii k \lambda_j}/(e^{\ii \pi \nu_j/2}\sqrt{|\det \mathcal{H}f(\bm{x}_j)|})$, and we accordingly set $\sqrt{\det \mathcal{H}f(\bm{x}_j)} = e^{\ii\pi\nu_j/2}\sqrt{|\det \mathcal{H}f(\bm{x}_j)|}$. With this convention, every real saddle point enters the Picard-Lefschetz formula with $n_j = 1$. For a negative determinant, the principal branch is ambiguous, so this choice has to be made explicitly. Only the combinations $n_j/\sqrt{\det\mathcal{H}f(\bm{x}_j)}$ are independent of the convention, and the same holds for the Stokes constants introduced in \cref{sec:Borel}.

    \paragraph{The transseries.}
    Each thimble integral admits an asymptotic expansion in inverse powers of $k$. The Picard-Lefschetz formula then yields the transseries
    \begin{align}
        \Psi \sim \sum_{j=1}^n n_j\, e^{\ii k \lambda_j} \sum_{m=0}^\infty \frac{T_m^{(j)}}{k^m}\,,
    \end{align}
    with $T_0^{(j)} = 1/\sqrt{\det \mathcal{H}f(\bm{x}_j)}$. The contribution of a complex saddle point is suppressed by the factor $|e^{\ii k \lambda_j}| = e^{-k\,\mathrm{Im}\,\lambda_j}$. As a saddle point is relevant when its ascent manifold intersects the original integration domain, along which $h(\bm{x}) = 0$, the relevant complex saddle points are generally suppressed, \textit{i.e.}, $\text{Im } \lambda_j \geq 0$. The asymptotic series generically diverge, and their resummation is the subject of the remainder of this paper.

    \paragraph{Stokes phenomena and the problem of relevance.}
    The thimbles, and with them the intersection numbers, depend on the phase of $k$ and on any parameters in the exponent. As long as no path of steepest descent connects two saddle points, the thimbles deform continuously and the intersection numbers remain constant. When the steepest descent manifold of $\bm{x}_j$ runs into a saddle point $\bm{x}_l$, which requires the phases to coincide, $\mathrm{Im}[\ii k \lambda_j] = \mathrm{Im}[\ii k \lambda_l]$, the thimble $\mathcal{J}_j$ jumps by a multiple of $\mathcal{J}_l$. Across such a Stokes phenomenon, the intersection numbers change, while the integral $\Psi$ itself remains continuous. Two saddle points that undergo a Stokes phenomenon for some phase of $k$ are called adjacent, and the integer jump of the thimble is the Stokes constant.

    The intersection numbers $n_j$ thus encode the global structure of the integral, and they are the first obstacle to any asymptotic evaluation. Traditionally, they are evaluated by integrating the gradient flow in $\mathbb{C}^d$ and counting the intersections of the steepest ascent manifolds with $\mathbb{R}^d$. In one dimension, this amounts to plotting the steepest descent contours. In higher dimensions, the flow lives in $2d$ real dimensions, and the computation rapidly becomes intractable \cite{Kaminsk:i1994, Shoji:2026}. In this paper, we instead obtain the intersection numbers algebraically. The Gauss-Manin system (\cref{sec:GaussManin}) yields the asymptotic series of all saddle points (\cref{sec:asymptotic}) and, through the Fuchsian system in the Borel plane, the Stokes constants (\cref{sec:Borel}). The South-East rule then converts the adjacency into the intersection numbers (\cref{sec:approximation}).

    \section{The Gauss-Manin system}\label{sec:GaussManin}
    Given a saddle point $\bm{x}_j$, we develop the asymptotic series coefficients $T_m^{(j)}$ of the thimble integral $\Psi^{(j)}$ in \cref{eq:thimble} using a Gauss-Manin system. We follow the twisted de Rham cohomology \cite{Schwarz:2009} and Brieskorn formalism \cite{Brieskorn:1970} (for details see \cref{ap:twisted}). This analysis applies to general complex valued polynomial Morse functions $f \in \mathbb{C}[\bm{x}]$.

    As the integrand decays exponentially on the descent manifold $\mathcal{J}_j$ away from the saddle point $\bm{x}_j$, the integral over the derivative $\frac{\partial}{\partial x_i} [h(\bm{x}) e^{\mathrm{i} k f(\bm{x})}]$ vanishes by the divergence theorem, \textit{i.e.},
    \begin{align}
        \int_{\mathcal{J}_j} \left[\partial_j h(\bm{x}) + \mathrm{i} k h(\bm{x})\partial_j f(\bm{x})\right] e^{\mathrm{i} k f(\bm{x})}\mathrm{d}\bm{x} 
        = 0\,,
    \end{align}
    for any polynomial $h \in \mathbb{C}[\bm{x}]$.
    Given a moment integral, sometimes known as a period, 
    \begin{align}
        \langle g(\bm{x})\rangle = \int_{\mathcal{J}_j} g(\bm{x}) e^{\mathrm{i} k f(\bm{x})}\mathrm{d}\bm{x}\,,
    \end{align}
    with a general polynomial $g \in \mathbb{C}[\bm{x}]$, the integral is preserved under the substitution 
    \begin{align}
        g(\bm{x}) \mapsto g(\bm{x}) + [\partial_j h(\bm{x}) + \mathrm{i} k  h(\bm{x})\partial_j f(\bm{x})]\,.
    \end{align}
    
    With this relation, we can simplify the analysis of the moment. The polynomials $\mathbb{C}[\bm{x}]$ modulo this equivalence relation is known as twisted de Rham cohomology (see appendix \ref{ap:twisted} for more details). When the exponent $f$ has $n$ non-degenerate critical points, the quotient space is generated by $n$ basis functions. The same observation underlies the counting of master integrals of Feynman integrals by the critical points of the Lee-Pomeransky polynomial, or the Euler characteristic of the associated variety \cite{Lee:2013hzt, Bitoun:2017nre, Agostini:2022cgv}. We denote the basis functions by $\bm{\phi}(\bm{x}) =(\phi_1(\bm{x}), \dots, \phi_n(\bm{x})) \in \mathbb{C}[\bm{x}]^n$. In the quotient algebra, the normal form of the polynomial $g(\bm{x})$ consists of a linear combination of basis functions, $g(\bm{x}) \equiv \bm{\alpha}(k) \cdot \bm{\phi}(\bm{x})$, with the vector-valued function $\bm{\alpha}(k) = (\alpha_1(k), \dots, \alpha_n(k))$. We identify the function $\bm{\alpha}(k)$ with Brieskorn reduction \cite{Brieskorn:1970}. Reduce $g(\bm{x})$ with respect to the basis $\bm{\phi}(\bm{x})$,
    \begin{align}
        g(\bm{x}) = \bm{m} \cdot \bm{\phi}(\bm{x}) + \bm{\beta}(\bm{x}) \cdot \nabla f(\bm{x})\,,
    \end{align}
    for some vector field $\bm{\beta}(\bm{x})$. Applying the equivalence relation, $h(\bm{x}) \partial_j f(\bm{x}) \mapsto -\frac{1}{\mathrm{i}k} \partial_j h(\bm{x})$ , we obtain the identity
    \begin{align}
        g(\bm{x}) \equiv \bm{m} \cdot \bm{\phi}(\bm{x}) -\frac{1}{\mathrm{i} k} \nabla \cdot \bm{\beta}(\bm{x})\,.
    \end{align}
    Repeating this process for the divergence $\nabla \cdot \bm{\beta}(\bm{x})$ yields a power series in $k^{-1}$. Consequently, the moment integral can be expressed as
    \begin{align}
        \langle g(\bm{x}) \rangle  = \bm{\alpha}(k) \cdot \bm{I}^{(j)}(k)\,,
    \end{align}
    with $\bm{I}^{(j)} = (I_1^{(j)}, \dots, I_n^{(j)})$ the $n$ standard moments,
    \begin{align}
        I_i^{(j)}(k) = \left(\frac{k}{2\pi \ii}\right)^{d/2}\int_{\mathcal{J}_j} \phi_i(\bm{x}) e^{\mathrm{i} k f(\bm{x})} \mathrm{d}\bm{x}\,.
    \end{align}
    For convenience, we let the first basis function be unity, $\phi_1(\bm{x})=1$, such that the first standard moment $I_1^{(j)}$ reduces to the original integral $\Psi^{(j)}$.

    \bigskip
    Differentiating the standard moments with respect to $k$, we obtain
    \begin{align}
        \frac{\mathrm{d}I_i^{(j)}(k)}{\mathrm{d} k} = \mathrm{i} \left(\frac{k}{2\pi \mathrm{i}}\right)^{d/2} \int_{\mathcal{J}_j} f(\bm{x})\phi_i(\bm{x}) e^{\mathrm{i} k f(\bm{x})}\mathrm{d}\bm{x}
        + \frac{d}{2 k} I_i^{(j)}(k)
        \,.
    \end{align}
    Replacing the product $f(\bm{x}) \phi_i(\bm{x})$ by its normal form $\bm{\alpha}_i(k) \cdot \bm{\phi}(\bm{x})$, we obtain the differential equation
    \begin{align}
        \frac{\mathrm{d} \bm{I}^{(j)}(k) }{\mathrm{d}k } = \left[\mathrm{i} M  
        - \sum_{r=1}^\infty \frac{B_{r}}{\mathrm{i}^{r-1} k^r}
        + \frac{d}{2k}I\right] \bm{I}^{(j)}(k)\,, 
    \end{align}
    with the matrices $M$ and $B_r$ determined by the normal forms,
    \begin{align}
        [\bm{\alpha}_i(k)]_j = M_{ij} - \sum_{r=1}^\infty \frac{[B_r]_{ij}}{ (\mathrm{i} k)^r}\,.
    \end{align}
    Whether the Brieskorn reduction process terminates, and if so, at what point, is determined on the choice of basis $\bm{\phi}$. Remarkably, Saito showed that, for polynomials with no critical points at infinity, we can select a \textit{good basis} $\bm{\phi}(\bm{x})$ for which the process terminate after one step \cite{Saito:1989, Saito:2013}, yielding the Gauss-Manin system 
    \begin{align}
        \frac{\mathrm{d} \bm{I}^{(j)}(k) }{\mathrm{d}k } = \left[\mathrm{i} M  
        + \frac{1}{k}\left(\frac{d}{2} I  - B\right)
        \right] \bm{I}^{(j)}(k)\,.\label{eq:GaussManin}
    \end{align}
    This differential equation does not depend on the details of the descent contour, as the matrices $M$ and $B$ are independent of the thimble $\mathcal{J}_j$. So will the original integral, as the Picard-Lefschetz formula expresses it as a linear combination of the integrals $\Psi^{(j)}$ along the descent manifolds $\mathcal{J}^{(j)}$. Remarkably, integration by parts has allowed us to transform the resurgent study of any exponential multidimensional integral with a polynomial exponent into the resurgent study of Gauss-Manin systems.
    
    The matrix $M$ represents the linear operation of multiplication by $f$ in the Jacobi algebra of $f$. By Stickelberger's theorem, the eigenvalues $\lambda_j$ of the matrix $M$ coincide with the critical values $f(\bm{x}_j)$, \textit{i.e.},
    \begin{align}
        \lambda_j = f(\bm{x}_j)\,.
    \end{align}
    The associated eigenvectors assume the form 
    \begin{align}
        \bm{v}_j = \frac{\bm{\phi}(\bm{x}_j)}{\sqrt{\det \mathcal{H} f(\bm{x}_j)}}\,.
    \end{align}

    In this paper, we derive the Gauss-Manin system with \textit{good} basis functions with the \verb|gmspoly.lib| library of the \verb|Singular| programming language \cite{Sabbah:1999, Schulze:2003}. In particular, we use the \verb|Singular.jl| implementation in the \verb|Julia| programming language.

    \subsection{The Brieskorn reduction of the Airy integral}
    We illustrate the derivation of the Gauss-Manin system for the Airy integral associated with the fold catastrophe
    \begin{align}
        \Psi(x) = \sqrt{\frac{k}{2\pi i}} \int e^{\mathrm{i} k (u^3 + x u)}\mathrm{d}u\,,
    \end{align}
    for the external parameter $x$.
    The quotient space is defined by the identity
    \begin{align}
        h'(u) +  \mathrm{i} k h(u) f'(u) \equiv 0\,,
    \end{align}
    with $f=u^3 + x u$.
    For the monomials $h(u) = 1,u$ and $u^2$, we obtain the reduction identities
    \begin{align}
        u^2 \equiv - \frac{x}{3}\,, \quad
        u^3 \equiv - \frac{x u}{3} - \frac{1}{3 \mathrm{i} k} \,, \text{ and }
        u^4 \equiv   \frac{x^2}{9} - \frac{2u}{3 \mathrm{i} k} \,. 
    \end{align}
    Every monomial $u^n$ reduces to a normal form consisting of a linear combination of the basis elements $\bm{\phi}(u) = (1,u)$. Reducing the products $\phi_i(u) f(u)$ yields 
    \begin{align}
        \phi_1(u) f(u) \equiv   \frac{2 x u}{3} - \frac{1}{3 \mathrm{i} k} \,,\quad 
        \phi_2(u) f(u) \equiv - \frac{2x^2}{9}  - \frac{2u}{3 \mathrm{i} k} \,,
    \end{align}
    from which we obtain the matrices
    \begin{align}
        M = \frac{1}{9} \begin{pmatrix}
            0 & 6 x\\
            -2 x^2 & 0
        \end{pmatrix}
        \text{ and }
        B = \frac{1}{3}\begin{pmatrix}
            1 & 0 \\
            0 & 2
        \end{pmatrix}.
    \end{align}
    As the process terminates after one reduction step, the basis $\bm{\phi}(u) = (1,u)$ is known as a good basis \cite{Saito:1989}.
    
    \section{Asymptotic series}\label{sec:asymptotic}
    To study the properties of the integral $\Psi$, we develop a power series expansion for the period $\bm{I}^{(j)}$,
    \begin{align}
        \bm{I}^{(j)}(k) \sim e^{\mathrm{i} k  \lambda_j} (\mathrm{i}k)^{\mu_j} \sum_{m=0}^\infty \frac{\bm{T}_m^{(j)}}{k^m}\,, \label{eq:ansatz}
    \end{align}
    with the coefficients $\lambda_j$, $\mu_j$, and $\bm{T}_m^{(j)}$. Substituting this ansatz in the Gauss-Manin equation \eqref{eq:GaussManin} and sorting in inverse powers of $k$ yields system of equations. The zeroth order in $k$, the Gauss-Manin equation reduces to the eigenequation
    \begin{align}
        M  \bm{T}_{0}^{(j)} = \lambda_j \bm{T}_0^{(j)}\,.
    \end{align}
    The coefficient $\lambda_j$ is an eigenvalue of the matrix $M$. As we discussed in the previous section, each eigenvalue $\lambda_j$ corresponds to the critical value $f(\bm{x}_j)$ of the critical point $\bm{x}_j$. The vector $\bm{T}_0^{(j)}$ is the associated eigenvector, normalized such that the first row 
    \begin{align}
        \left[\bm{T}^{(j)}_0\right]_{1}=\frac{1}{\sqrt{\det \mathcal{H} P(\bm{x}_j)}}
    \end{align}
     for our choice of bases $\phi_1(\bm{x})=1$. The higher-order coefficients $\bm{T}_m^{(j)}$ follow through Frobenius recursion from the higher-order inverse powers of $k$ \cite{Nemes:2013:GFSN}. We explicitly derive the recursion relations in appendix \ref{ap:ODE} leading to  \cref{alg:asymptoticTerms}. Summing over the eigenvalues (or equivalently the saddle points), we recover the transseries expansion of the periods
    \begin{align}
        \bm{I}(k) \sim \sum_j n_j e^{\mathrm{i} k \lambda_j} (\mathrm{i}k)^{\mu_j} \sum_{m=0}^\infty \frac{\bm{T}_m^{(j)}}{k^m}\label{eq:transseries}\,.
    \end{align}
    The first component of $\bm{I}(k)$ is the transseries expansion of our integral of interest. For non-degenerate critical points, we have defined the normalization of the origin integral such that  $\mu_j=0$.

    \begin{algorithm}[H]
        \caption{Algorithm for the evaluation of the coefficients $\bm{T}_m^{(j)}$ of the asymptotic series.}
        \label{alg:asymptoticTerms}
        \begin{algorithmic}[1]
            \State Compute a good basis $\bm{\phi}$ and the associated Gauss-Manin system with the matrices $M$ and $B$.
            \State Evaluate the eigenvalues $\lambda_j$ and associated eigenvectors $\bm{v}_j$ of the matrix $M$, forming the diagonal matrix 
            \begin{align}
                \Lambda=\text{diag}(\lambda_1,\dots,\lambda_n)
            \end{align}
            and eigenvector matrix 
            \begin{align}
                V=(\bm{v}_1,\dots,\bm{v}_n)\,.
            \end{align} 
            \State Construct the matrices 
            \begin{align}
                \tilde{B}=\frac{d}{2}I - V^{-1}BV    
            \end{align}
            and extract the diagonal elements $\mu_j = \tilde{B}_{jj}$. These vanish for non-degenerate critical points.
            \State For a given index $j$, find the corresponding saddle point $\bm{x}_j$ and build the initial vector $\bm{Y}_0^{(j)} = \bm{e}_j$.
            \State Update all rows except for the $j$th row using 
                \begin{align}
                    Y_{i,m}^{(j)} = \frac{1}{\lambda_j - \lambda_i}\left( \sum_{l\neq i}\tilde{B}_{il} Y_{l,m-1}^{(j)} +( \mu_i - \mu_j + m - 1) Y_{i, m-1}^{(j)} \right)\,.
                \end{align}
            \State Update the $j$th row using
                    \begin{align}
                        Y_{j, m}^{(j)} = -\frac{1}{m}\sum_{l \neq j} \tilde{B}_{jl}Y_{l,m}^{(j)} \,.
                    \end{align}
            \State Repeat step 5 and 6 till we have evaluated the vectors $\bm{Y}_0^{(j)}, \bm{Y}_1^{(j)}, \dots,\bm{Y}_M^{(j)}$.
            \State Rotate the vectors back to obtain the coefficients $\bm{T}_m^{(j)}$ with $\bm{T}_m^{(j)} =  \mathrm{i}^{-m} \mathcal{N} V \bm{Y}_m^{(j)}$. In this identity, we select the normalization $\mathcal{N}$ such that $[\bm{T}^{(j)}_{0}]_1= 1/\sqrt{\det \mathcal{H}f(\bm{x}_j)}$.
        \end{algorithmic}
    \end{algorithm}

    \section{Into the Borel plane}\label{sec:Borel}
    The asymptotic series of the previous section diverge: the coefficients $\bm{T}_m^{(j)}$ grow factorially with the order $m$. The Borel transform tames this growth by dividing the $m$th coefficient by a factor $\Gamma(m + \cdot)$, which yields a function with a finite radius of convergence. The Laplace transform of this function then reproduces the original integral \cite{Ecalle:1981, Dingle:1973, sauzin2014intro, Aniceto:2018bis}. For oscillatory integrals, the Borel plane has a direct geometric meaning, which we briefly review following Pham and Howls \cite{Pham1965FormulesDP, Howls:1997}.

    Consider the thimble integral $\Psi^{(j)}$ of \cref{eq:thimble} and change the integration variable to the Borel variable
    \begin{align}
        t = -\ii f(\bm{x})\,,
    \end{align}
    such that the integrand becomes $e^{\ii k f(\bm{x})} = e^{-kt}$. As the phase $\mathrm{Im}[\ii k f]$ is constant along the thimble $\mathcal{J}_j$ and the magnitude decays away from the saddle point, the thimble maps to the ray in the Borel plane running from the critical value $t_j = -\ii\lambda_j$ to infinity in the direction $e^{-\ii\theta}$, with $\theta = \arg k$. Integrating over the level sets of $f$ first, we obtain the Laplace integral
    \begin{align}
        \Psi^{(j)}(k) = \left(\frac{k}{2\pi\ii}\right)^{d/2}\int_{t_j}^{t_j + e^{-\ii\theta}\infty} e^{-kt}\, \varphi_j(t)\,\dd t\,, \quad
        \varphi_j(t) = \int_{\mathcal{J}_j \cap \{-\ii f = t\}} \frac{\dd\bm{x}}{\dd(-\ii f)}\,,
    \end{align}
    where $\varphi_j(t)$ is the integral of the Gelfand-Leray form $\dd\bm{x}/\dd(-\ii f)$ over the $(d-1)$-dimensional intersection of the thimble with the level set $-\ii f = t$. The $d$-dimensional integral is reduced to a Laplace integral along a ray in a single complex plane, irrespective of the dimension $d$ \cite{Howls:1997, Aniceto:2026SouthEast}. The function $\varphi_j(t)$ is analytic away from the critical values $t_j$. Continuing $\varphi_j$ along a path in the Borel plane, the vanishing cycle of $\bm{x}_j$ may get pinched at another critical value $t_l$. When this happens, the two branch points $t_j$ and $t_k$ ``see each other'' on the Riemann surface representing the Borel transform $\varphi$ and are known as adjacent. The Borel plane thus turns the global resurgent structure of the integral into geometry \cite{aniceto2026algebraic}.

    \subsection{The Borel transform}\label{sec:BorelTransform}
    Define the Laplace $\widehat{\bm{I}}(t)$ by the identity
    \begin{align}
        k^{c-\frac{d}{2}} \bm{I}^{(j)}(k) = \int_{t_j}^{t_j+e^{-\ii\theta}\infty} e^{-k t} \widehat{\bm{I}}^{(j)}(t) \mathrm{d}t\,,
        \label{eq:Laplace}
    \end{align}
    running from the point $t_j = - \ii \lambda_j$ to infinity $e^{-\ii \theta} \infty$ with the angle $\theta = \arg k$ ensuring the exponential decays without any oscillations, mirroring the integrand of the origin integral along the steepest descent manifold $\mathcal{J}_j$. The constant $c$ aid the Borel analysis, but while not significantly altering the resurgent structure of the integral. Upon substituting the expansion 
    \begin{align}
        \widehat{\bm{I}}^{(j)}(t) = (t-t_j)^{\rho}\sum_{m =0}^\infty \bm{c}_m \,(t-t_j)^m\,,\label{eq:laplaceExpansion}
    \end{align}
    in the Laplace transform, and matching terms of the expansion of $\bm{I}^{(j)}$, we obtain the identification $\rho = \frac{d}{2} - 1 -\mu_j - c$ and 
    \begin{align}
        \bm{c}_m^{(j)} = \frac{\ii^{\mu_j}\bm{T}_m^{(j)}}{\Gamma(1+ m + \rho)}\,.
    \end{align}
    For non-degenerate critical points $\mu_j = 0$, we obtain the famous Borel transform of the asymptotic series,
    \begin{align}
        \widehat{\bm{I}}^{(j)}(t)
         = (t-t_j)^{\frac{d}{2}-1 - c} \sum_{m=0}^\infty \frac{\bm{T}_m^{(j)} (t-t_j)^m}{\Gamma(m + \frac{d}{2}  -c)}\,.
    \end{align}

    The nature of the series expansion changes for different constants $c$:
    \begin{itemize}
        \item For $c= (d-1)/2$, the saddle points $\bm{x}_j$ map to square root branch point singularities at $t_j$ as $\rho = -1/2$. The Laplace transform assume the form 
        \begin{align}
            \widehat{\bm{I}}^{(j)}(t)
             = (t-t_j)^{-1/2} \sum_{m=0}^\infty \frac{\bm{T}_m^{(j)} (t-t_j)^m}{\Gamma(m + \frac{1}{2} )}\,.
        \end{align}
        \item For $c = \frac{d}{2}-1$, the overall scaling is removed $\rho =0$ and we obtain the simplest Borel transform 
        \begin{align}
            \widehat{\bm{I}}^{(j)}(t)
             = \sum_{m=0}^\infty \frac{\bm{T}_m^{(j)} (t-t_j)^m}{m!}\,.
        \end{align}
        \item For the choice $c = \frac{d}{2}$, without the modified scaling law, the Borel transform scales as $(t-t_j)^{-1}$. The gamma function diverges for the term $m=0$. Consequently, we would need to remove $\bm{T}_0^{(j)}$ before performing the resurgence analysis.
    \end{itemize}
    Contrary to the asymptotic series for $\bm{I}^{(j)}(k)$, the Borel transform $\widehat{\bm{I}}^{(j)}(t)$ converges on the disk centred at $t_j$ reaching the nearest other singularity. This is known as the Frobenius disk.
    
    \begin{figure}
        \centering
        \includegraphics[width = 0.75 \textwidth]{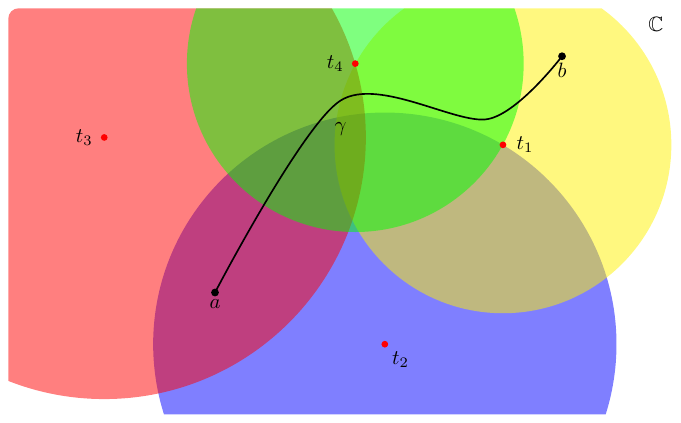}
        \caption{Analytic continuation in the Borel plane. The red points are the singular points $t_j = -\ii\lambda_j$ of the Fuchsian system. Around each singular point, the local Frobenius expansion of the solution converges on a disk reaching the nearest other singular point (coloured disks). The propagator $U_\gamma$ continues a solution from $a$ to $b$ along a path $\gamma$ that avoids the singular points. The result depends only on the homotopy class of $\gamma$ in $\mathbb{C}\setminus\{t_1,\dots,t_n\}$, \textit{i.e.}, on how $\gamma$ winds around the singular points.}\label{fig:BorelTransform}
    \end{figure}

    \bigskip
    The Laplace transform exchanges the parameter $k$ with the Borel parameter $t$, mapping the Gauss-Manin equation \eqref{eq:GaussManin} to the Fuchsian system
    \begin{align}
        \frac{\mathrm{d}\widehat{\bm{I}}^{(j)}(t)}{\mathrm{d} t} = A(t) \widehat{\bm{I}}^{(j)}(t)\,,
    \end{align}
    with the matrix 
    \begin{align}
        A(t) = \left[t I + \mathrm{i}M \right]^{-1}\left[ B - (1 + c) I \right]\,.
    \end{align}
    The matrix $A(t)$ is independent of the saddle point of interest and has $n$ singularities located at the branch points $t_j$ in the Borel plane associated with the critical points $\bm{x}_j$. 
    The Fuchsian system is a set of linear ordinary differential equations, whose solutions can be analytically continued along any path that avoids the singularities $t_j$. The Borel transforms $\widehat{\bm{I}}^{(j)}$ are resurgent: they live on the universal cover of $\mathbb C \setminus\{t_1,\dots,t_n\}$ and have no other singularities. The expansion $\widehat{\bm{I}}^{(j)}(t)$ around the singularities $t_j$ can be interpreted as local expansions of the Riemann surface $\widehat{\bm{I}}(t)$. Explicitly, the Fuchsian system is solved by the path-ordered exponential
    \begin{align}
        \widehat{\bm{I}}(t_1) = \mathcal{P} e^{ \int_{\gamma} A(t)\mathrm{d}t} \widehat{\bm{I}}(t_0)\,,
    \end{align}
    where $\gamma \subset \mathbb{C}$ is a path in the Borel plane moving from $t_0$ to $t_1 \in \mathbb{C}$.
    The propagator $U(t_1,t_0)=\mathcal{P} e^{ \int_\gamma A(t)\mathrm{d}t}$ continues the vector $\widehat{\bm{I}}(t_0)$ to $\widehat{\bm{I}}(t_1)$ along a path $\gamma$ in the Borel plane. The value $\widehat{\bm{I}}(t_1)$ depends only on the homology of the path $\gamma$ in the universal covering of $\mathbb{C} \setminus \{t_1,\dots, t_n\}$, \textit{i.e.} deforming $\gamma$ without crossing the singularities $t_j$ preserves the Borel transform $\widehat{\bm{I}}(t_1)$. See \cref{fig:BorelTransform} for an illustration of the Borel plane, the Frobenius disks and the path $\gamma$.


    \begin{figure}
        \centering
        \begin{subfigure}{0.48 \textwidth}
            \includegraphics[width=\textwidth]{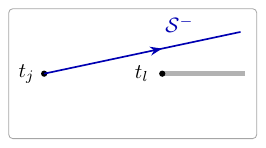}
            \caption{}
        \end{subfigure}
            \begin{subfigure}{0.48 \textwidth}
            \includegraphics[width=\textwidth]{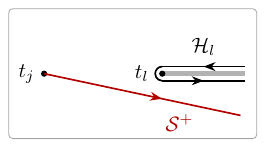}
            \caption{}
        \end{subfigure}
        \caption{The Stokes phenomenon in the Borel plane. (a) Before the transition, the Laplace ray $\mathcal{S}^-$ of the saddle point $\bm{x}_j$ runs from $t_j$ above the singularity $t_l$ of an adjacent saddle point. (b) As $\theta = \arg k$ increases, the ray rotates clockwise and passes to the other side of $t_l$, becoming $\mathcal{S}^+$. The integrals along the two rays differ by $K_{lj}$ times the integral along the Hankel contour $\mathcal{H}_l$ around the branch cut emanating from $t_l$ (grey), which equals $\bm{I}^{(l)}(k)$, such that $\mathcal{S}^+\bm{I}^{(j)} - \mathcal{S}^-\bm{I}^{(j)} = K_{lj}\bm{I}^{(l)}$. In terms of the thimbles, $\mathcal{J}_j$ picks up $K_{lj}$ copies of $\mathcal{J}_l$.}\label{fig:StokesTransition}
    \end{figure}

    \begin{figure}
        \centering
        \includegraphics[width = \textwidth]{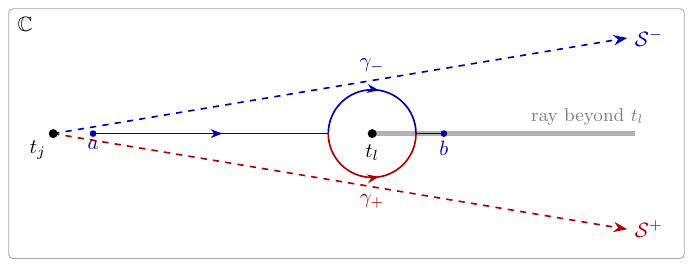}
        \caption{The keyhole contours used to evaluate the Stokes constant $K_{lj}$. The paths start at $a$, in the Frobenius disk of $t_j$, run straight towards $t_l$, pass $t_l$ on a half-circle above ($\gamma_-$, blue) or below ($\gamma_+$, red), and end at $b$, in the Frobenius disk of $t_l$. They correspond to the Laplace rays $\mathcal{S}^-$ and $\mathcal{S}^+$ (dashed) before and after the Stokes transition, deformed towards the line through $t_j$ and $t_l$. The difference of the propagators along the two paths yields the discontinuity $\bm{\Delta}(b) = (U_+ - U_-)\,\widehat{\bm{I}}^{(j)}(a) = K_{lj}\,\widehat{\bm{I}}^{(l)}(b)$ across the branch cut beyond $t_l$ (grey).}\label{fig:keyhole}
    \end{figure}

    \subsection{Evaluating Stokes constants}\label{sec:StokesConstants}
    Historically, the adjacency relations were obtained from the asymptotic series using either the hyperasymptotic expansion \cite{Daalhuis:1999} or using Borel-Pad\'e methods (for a brief summary and references see \cite{Aniceto:2018bis,lustri2023exponential}). In this section, we propose to evaluate the adjacency relations directly from the Fuchsian system following Ecalle \cite{Ecalle:1981,Ecalle:1993}.

    When rotating the parameter $k = |k|e^{\ii \theta}$ around the origin in the complex plane, the integration contour of the Laplace transform \cref{eq:Laplace}, known as the Laplace ray $(t_j,  t_j + e^{-\ii \theta}\infty)$ of $\bm{I}^{(j)}(k)$, rotates in the Borel plane. 
    As it encounters another branch point, $t_l$ on the same Riemann sheet of the Borel transform $\widehat{\bm{I}}(k)$, the integral undergoes a Stokes transition. After the Stokes transition, the integral acquires a contribution from the integral $\bm{I}^{(l)}(k)$ (see \cref{fig:StokesTransition}). When such a transition occurs, between saddle points $\bm{x}_j$ and $\bm{x}_l$ for some angle $\theta \in [0, 2\pi)$, the two saddle points are called adjacent. Given such a Stokes transition, we consider the integral along the Laplace ray before $\mathcal{S}^-\bm{I}^{(j)}(k)$ and after $\mathcal{S}^-\bm{I}^{(j)}(k)$ the Stokes transition (see \cref{fig:keyhole}). Their difference is given by 
    \begin{align}
        \mathcal{S}^+\bm{I}^{(j)}(k) - \mathcal{S}^-\bm{I}^{(j)}(k) = K_{lj} \bm{I}^{(l)}(k)\,,
    \end{align}
    with the Stokes constant $K_{lj} \in \mathbb{Z}$. During the Stokes transition, the thimble $\mathcal{J}_j$ picks up $K_{lj}$ copies of thimble $\mathcal{J}_l$. 
    
    Now, deform both rays towards the line through $t_j$ and $t_l$. On the segment from $t_j$ to $t_l$ their continuation of the Borel transforms $\widehat{\bm{I}}^{(j)}(t)$ coincide. The small arcs around $t_l$ do not contribute, as the singularity at $t_l$ is integrable. What remains is the ray beyond $t_l$, traversed with the continuation $\widehat{\bm{I}}^{(j)}_+(t)$ below and $\widehat{\bm{I}}^{(j)}_-(t)$ above the point $t_l$,
    \begin{align}
        \mathcal{S}^+\bm{I}^{(j)}(k) - \mathcal{S}^-\bm{I}^{(j)}(k) = \int_{t_l}^{t_l + e^{-\ii \theta} \infty} e^{- k t} \bm{\Delta}(t)\mathrm{d}t\,,
    \end{align}
    with the difference 
    \begin{align}
        \bm{\Delta}(t) = \widehat{\bm{I}}^{(j)}_+(t) - \widehat{\bm{I}}^{(j)}_-(t)\,.
    \end{align}
    Together, the two contours form a Hankel contour around the ray from $t_l$. As the Laplace integrals coincide with $K_{lj} \bm{I}^{(l)}(k)$ and the difference $\bm{\Delta}$ is a solution of the Fuchsian system (being the difference of two solutions), we obtain the relation
    \begin{align}
        \bm{\Delta}(t) = K_{lj} \widehat{\bm{I}}^{(l)}(t)\,.
    \end{align}
    The fact that the number $K_{lj}$ is an integer follows from the intersection numbers of the Picard-Lefschetz formula. Moreover, as $\widehat{\bm{I}}^{(j)}_{+}(t)$ is obtained from $\widehat{\bm{I}}^{(j)}_{-}(t)$ by one loop around $t_l$, we obtain $\widehat{\bm{I}}^{(j)}_{+}(t) = \mathcal M_l^{\pm1}\widehat{\bm{I}}^{(j)}_{-}(t)$ with $\mathcal{M}^{\pm 1}$, leading to the identity
    \begin{align}
        \bm{\Delta}(t) = (\mathcal{M}_l^{\pm 1} - I)\widehat{\bm{I}}^{(j)}_-(t)\,,
    \end{align}
    relating the Stokes constants to the monodromy of the Fuchsian system. 
    
    At first sight, it may seem we need to continue the difference $\bm{\Delta}(t)$ along the Laplace ray emanating from $t_l$. However, as the difference of two solutions of the Fuchsian system, it is itself a solution. As the analytic continuation is completely determined by its value in a point in the Borel plane, it suffices to compare the difference $\bm{\Delta}(b)$ and the Borel transform $\bm{I}^{(l)}(b)$ at a point $b$ in the Frobenius disk of $t_l$. This allows us to evaluate the Stokes constants using the local expansions of the Borel transforms and linear algebra along two paths.
    \begin{enumerate}
        \item \textbf{Paths.} Choose points $a$ and $b$ in the Frobenius disks of $t_j$ and $t_l$. The path $\gamma_\pm$ run from $a$ straight towards $t_l$, pass it on a half-circle above ($-$) or below ($+$), and end at $b$.
        \item \textbf{Propagators.} Compute the propagators $U_\pm = \mathcal P e^{\int_{\gamma_\pm}A(t)}\,\dd t$ along the two paths. We can do this either using Taylor steps or hyperlogarithms (see the next section).
        \item \textbf{Saddle solutions.} Evaluate $\widehat{\bm{I}}^{(j)}(a)$ and $\widehat{\bm{I}}^{(l)}(b)$ from the expansion \eqref{eq:laplaceExpansion}, with $\bm{c}_0 = T_0\,\bm{v}/v_1$ (the eigenvector with unit first entry times $T_0 = 1/\sqrt{\det\mathcal H f}$ of the corresponding saddle).
        \item \textbf{Stokes constant.} Form the jump $\bm\Delta(b) = (U_+-U_-)\,\widehat{\bm{I}}^{(j)}(a)$ and evaluate the Stokes constant
        \begin{equation}
            K_{lj} = \frac{\widehat{\bm{I}}^{(l)}(b)^{\dagger}\bm\Delta(b)}{\widehat{\bm{I}}^{(l)}(b)^{\dagger}\widehat{\bm{I}}^{(l)}(b)} .
        \end{equation}
        When the calculation has converged, the residual $\|\bm\Delta - K_{lj} \widehat{\bm{I}}^{(l)}(b)\|$ should be sufficiently small.
    \end{enumerate}

    \paragraph{Single-loop formula.}
    The difference $U_+ - U_-$ of two nearly equal propagators is prone to cancellation, in particular when the propagators are large compared to the Stokes constant. The subtraction can be avoided by noting that the paths $\gamma_\pm$ share the stem from $a$ to the point $p$ in the Frobenius disk of $t_l$ where they leave the line through $t_j$ and $t_l$, and differ only in the half-circles around $t_l$. The half-circle below $t_l$ is homotopic to a counterclockwise loop $\ell_l$ around $t_l$ starting at $p$, followed by the half-circle above $t_l$. As the propagators compose along paths, $U_+ = U_{\mathrm{arc}} L_l U_{\mathrm{stem}}$ and $U_- = U_{\mathrm{arc}} U_{\mathrm{stem}}$, with the propagator $U_{\mathrm{stem}}$ from $a$ to $p$, the propagator $U_{\mathrm{arc}}$ along the upper half-circle from $p$ to $b$, and the local monodromy $L_l$ of the loop $\ell_l$. Consequently,
    \begin{align}
        \bm{\Delta}(b) = U_{\mathrm{arc}}\left(L_l - I\right) U_{\mathrm{stem}}\,\widehat{\bm{I}}^{(j)}(a)\,.
    \end{align}
    The local monodromy follows from the Frobenius normal form of the Fuchsian system at $t_l$. As the eigenvalues of the residue $R_l$ do not differ by positive integers, the solutions near $t_l$ assume the form $Y(t) = G_l(t)\,(t-t_l)^{R_l}\,C$, with a constant matrix $C$ and a matrix $G_l(t) = \sum_{m=0}^\infty G_m^{(l)} (t-t_l)^m$ that is holomorphic on the Frobenius disk, with $G_0^{(l)} = I$. Substituting this form in the Fuchsian system, with the expansion $A(t) = R_l/(t-t_l) + \sum_{m\geq 0} A_m^{(l)}(t-t_l)^m$, yields the recursion
    \begin{align}
        m\,G_m^{(l)} + G_m^{(l)} R_l - R_l\, G_m^{(l)} = \sum_{i=0}^{m-1} A_i^{(l)} G_{m-1-i}^{(l)}\,, \quad m \geq 1\,,
    \end{align}
    which has a unique solution as the eigenvalues of the commutator with $R_l$ are differences of eigenvalues of $R_l$. A counterclockwise loop multiplies $(t-t_l)^{R_l}$ by $e^{2\pi\ii R_l}$, such that $L_l = G_l(p)\, e^{2\pi\ii R_l} G_l(p)^{-1}$. For the choice $c = \frac{d}{2}-1$, the residues are nilpotent, $R_l^2 = 0$, and the exponential reduces to $e^{2\pi\ii R_l} = I + 2\pi\ii R_l$. We obtain the single-loop formula
    \begin{align}
        \bm{\Delta}(b) = 2\pi\ii\; U_{\mathrm{arc}}\, G_l(p)\, R_l\, G_l(p)^{-1}\, U_{\mathrm{stem}}\,\widehat{\bm{I}}^{(j)}(a)\,,\label{eq:singleLoop}
    \end{align}
    which evaluates the discontinuity directly as a product, without subtracting two transported solutions. As the residue $R_l = \Pi_l(B-(1+c)I)$ has rank one, the matrix $L_l - I$ projects any solution onto the saddle solution of $\bm{x}_l$, which is the Picard-Lefschetz formula for vanishing cycles in the Borel plane: the coefficient of this projection is the Stokes constant $K_{lj}$. Moreover, the Frobenius factor $G_l(t)$ and the residue $R_l$ depend only on the singularity $t_l$. They are computed once and reused for the Stokes constants $K_{lj}$ of all saddle points $\bm{x}_j$, with only the point $p$ at which $G_l$ is evaluated depending on the direction of the path from $t_j$.

    When the singularities $t_j$ and $t_l$ lay on the same Riemann sheets of the Riemann surface $\widehat{\bm{I}}(t)$ and they can see each other, they are adjacent, $K_{lj} \neq 0$. The adjacency materializes as the nontrivial monodromy of a loop starting in the vicinity of $t_j$ and circling $t_l$. When the singularities lay on different Riemann sheets, and they are hidden from each other, the monodromy of the loop starting at $t_j$ is trivial. The difference $\bm{\Delta}(t)$ vanishes, leading to the Stokes constant $K_{lj}=0$. 

    \subsection{Analytic continuation in the Borel plane}\label{sec:continuation}
    We evaluate the propagators $U_\pm$ along the contours $\gamma_\pm$ with either the Taylor step or Ecalle's mould method \cite{Ecalle:1981,Ecalle:1993}. Before explaining these methods in detail, we decompose the matrix $A(t)$ into its partial fractions 
    \begin{align}
        A(t) = \sum_{j=1}^n \frac{R_j}{t + \mathrm{i} \lambda_j} \,,
    \end{align}
    with the residue associated to the singularities
    \begin{align}
        R_j &= \Pi_j (B - \left(1+c\right)I)\,,
    \end{align} 
    defined in terms of the projection operator 
    \begin{align}
        \Pi_j = \prod_{i \neq j} \frac{M- \lambda_i I}{\lambda_j - \lambda_i}\,.
    \end{align} 
    This decomposition of $A(t)$ leads to a second route to evaluating the coefficients $\bm{c}_m$. Near $t_j$ write
    \begin{align}
        A(t) = \frac{R_j}{t-t_j} + \sum_{m=0}^\infty A_m^{(j)}\,(t-t_j)^m\,,
    \end{align}
    where the $A^{(j)}_m$ come from the other poles. Substituting the expansion \eqref{eq:laplaceExpansion}, we obtain the Frobenius recursion
    \begin{align}
      \left((\rho + n)I - R_j\right)\bm{c}_n^{(j)} = \sum_{m=0}^{n-1}A^{(j)}_m\,\bm{c}_{n - 1 - m}^{(j)}\,,
    \end{align}
    for $n \geq 1$. The matrix on the left-hand side is invertible, yielding the coefficients of the converging series expansion of the Borel transform around $t_j$.

    \subsubsection{Taylor steps}
    The partial fraction decomposition enables us to compute the propagators $U_\pm$ by decomposing the paths $\gamma_\pm$ into linear line segments. Each step is constructed to lie inside a disk where $A(t)$ is analytic. The algorithm follows three steps:
    \begin{enumerate}
        \item Construct a path $[p_1, \dots, p_N]$ with $p_1=a$ and $p_N=b$ in the equivalence class of $\gamma$, such that for consecutive points the second point lies in the Frobenius disk of the first point, \textit{i.e.}, 
        \begin{align}
            |p_{n+1} - p_n| < d(p_{n})
        \end{align}
        with $d(p)$ the distance of $p$ to the nearest singularity.
        \item  When moving from $p$ to $u$ in the Borel plane, expand the matrix $A(p+u) = \sum_{m = 0}^\infty A_m u^m$ with the coefficients $A_m = \sum_j \frac{R_j(-1)^m}{(p-t_s)^{m+1}}$. Upon substituting the series expansion of the propagator $U(p+u, p) = \sum_{m=0}^M C_m u^m$ truncated at $M$ into the Fuchsian equation $\frac{\mathrm{d}Y}{\mathrm{d}t} = AY$, we obtain the recursion relation
        \begin{align}
            C_{n+1} = \frac{1}{n+1}\sum_{m=0}^{n}A_m\,C_{n-m}\,,
        \end{align}
        with $C_0 = I$.
        \item The propagator $U$ is given by the matrix product of the intermediate steps 
        \begin{align}
            U = U(b, p_{N-1}) U(p_{N-1}, p_{N-2})\cdots U(p_1, a)\,.
        \end{align}
    \end{enumerate}

    \subsubsection{Ecalle's mould method}
    Alternatively, we may approximate the time-ordered exponential
    \begin{align}
        U(t_1,t_0)=\mathcal{P} e^{ g \int_{\gamma} A(t)\mathrm{d}t}
    \end{align}
    with the Dyson series,
    \begin{align}
        U(t_1,t_0) 
        =I 
        + g \int_{t_0 \leq s \leq t_1} A(s)\mathrm{d}s
        + g^2 \int_{t_0 \leq s_1 < s_2 \leq t_1} A(s_2) A(s_1)\mathrm{d}s_1\mathrm{d}s_2
        +
        \dots 
    \end{align}
    where $g$ is a bookkeeping parameter which we set to unity. With the partial fractions decomposition of $A(t)$, we obtain the identity
    \begin{align}
        U(t_1,t_0) 
        =I 
        + g \sum_{j=1}^n R_j \int_{t_0 \leq s \leq t_1} \frac{\mathrm{d}s}{s + \mathrm{i}\lambda_j}
        + g^2\sum_{j_1,j_2=1}^n R_{j_2} R_{j_1} \int_{t_0 \leq s_1 < s_2 \leq t_1} \frac{\mathrm{d}s_1\mathrm{d}s_2}{(s_1 + \mathrm{i}\lambda_{j_1})(s_2 + \mathrm{i}\lambda_{j_2})} 
        +
        \dots \,.
    \end{align}
    The integrals 
    \begin{align}
        L_{\sigma_1 \dots \sigma_r}(t) = \int_{0 \leq s_1 \leq \dots \leq s_r  \leq t} \frac{\mathrm{d}s_1 \dots \mathrm{d}s_r}{(s_1 - \sigma_1)\cdots (s_r  - \sigma_r)} \,,
    \end{align}
    are known as hyperlogarithms. The Riemann sheet structure of the Borel transform $\widehat{\bm{I}}$ is neatly enclosed in the residues $R_j$ and the multivalued nature of the hyperlogarithms. We decompose the path $\gamma$ into small steps $[p_1, p_2, \dots, p_N]$ with $p_1=a$ and $p_N=b$ and evaluate the propagator again using the product rule 
    \begin{align}
        U = U(b, p_{N-1}) U(p_{N-1}, p_{N-2})\cdots U(p_1, a)\,.
    \end{align}

    \bigskip

    \paragraph{Certified truncation.}
    The mould expansion may also be truncated with a rigorous stopping criterion. Let $U_\gamma^{(N)}$ denote the propagator obtained by retaining words of length at most $N$, and define
    \begin{align}
        \Lambda_\gamma
        =
        \int_\gamma \|A(t)\|\,|\dd t|.
    \end{align}
    The $r$th Dyson term is bounded by $\Lambda_\gamma^r/r!$, and hence
    \begin{align}
        \left\|U_\gamma-U_\gamma^{(N)}\right\|
        \leq
        E_U(N)
        =
        \sum_{r>N}\frac{\Lambda_\gamma^r}{r!}
        \leq
        e^{\Lambda_\gamma}
        \frac{\Lambda_\gamma^{N+1}}{(N+1)!}.
        \label{eq:DysonBound}
    \end{align}
    Thus the error from transport decreases factorially with the mould weight. For the choice $c=d/2-1$, for which $\rho=0$ and $R_j^2=0$, the local Frobenius solutions at the two ends of the keyhole also admit a super-exponentially decreasing remainder. Writing $z_s$ for the evaluation point near $t_s$, $\alpha_s=\|R_s\|$, and
    \begin{align}
        \lambda_s
        =
        |z_s-t_s|
        \max_{|t-t_s|\leq |z_s-t_s|}
        \left\|
            \sum_{u\neq s}\frac{R_u}{t-t_u}
        \right\|,
    \end{align}
    one may bound
    \begin{align}
        \|\delta\widehat{\bm I}^{(s)}\|
        \leq
        \|\bm v_s\| E_s(N),\qquad
        E_s(N)
        =
        \inf_{G>1}
        \frac{
            \exp\!\left[\lambda_s(G+\alpha_sG^2)\right]
        }{
            G^{N+1}(1-G^{-1})
        } .
        \label{eq:FrobeniusBound}
    \end{align}
    Combining \cref{eq:DysonBound,eq:FrobeniusBound} gives a computable bound
    \begin{align}
        |K_{ba}^{(N)}-K_{ba}|\leq B_{ba}(N),
        \label{eq:StokesBound}
    \end{align}
    including both transport and endpoint errors.\footnote{Explicitly, one may take
    \begin{align}
    B_{ba}(N)
    =
    \frac{
    2P_a E_U(N)
    +
    \|U_+-U_-\|\,\|\bm v_a\|E_a(N)
    +
    |K_{ba}|\,\|\bm v_b\|E_b(N)
    }{
    \|\widehat{\bm I}^{(b)}(b)\|
    -
    \|\bm v_b\|E_b(N)
    },
    \end{align}
    where $P_a\geq\|\widehat{\bm I}^{(a)}(a)\|$, provided the denominator is positive. All quantities may either be bounded a priori from the residues and path geometry, or evaluated a posteriori from a lower-cost computation.}

    The integrality of the Stokes constants now provides a particularly simple certificate. Since $K_{ba}\in\mathbb Z$, it is not necessary to take $N\to\infty$. As soon as
    \begin{align}
        B_{ba}(N)<\frac12,
        \label{eq:integerCertificate}
    \end{align}
    the error disk contains a unique integer and rounding $K_{ba}^{(N)}$ to the nearest integer recovers the exact Stokes constant. We may therefore stop at the first weight
    \begin{align}
        N_\ast
        =
        \min\left\{
            N:\,B_{ba}(N)<\frac12
        \right\}.
    \end{align}
    In this sense the mould expansion gives not only a convergent approximation to the Stokes matrix, but a finite certificate of its integer entries.

    \section{Approximating multidimensional oscillatory integrals}\label{sec:approximation}
    The previous sections provide, for any polynomial exponent $f$, three pieces of data: the critical values $\lambda_j$ (the eigenvalues of $M$), the coefficients $\bm{T}_m^{(j)}$ of the asymptotic series of every saddle point (\cref{alg:asymptoticTerms}), and the Stokes constants $K_{lj}$ encoding the adjacency of the saddle points. In this section, we outline how these data suffice to approximate the original integral $\Psi$ to any desired precision. The procedure consists of three steps. First, the South-East rule converts adjacency into relevance, \textit{i.e.}, determines the intersection numbers $n_j$ in the Picard-Lefschetz formula \cref{eq:PicardLefscehtz} \cite{Aniceto:2026SouthEast}. This step requires $f \in \mathbb{R}[\bm{x}]$. Second, optimally truncating the asymptotic series of the relevant saddle points yields an exponentially accurate superasymptotic approximation. Third, hyperasymptotics systematically improves on this approximation by resumming the remainders, again using only the adjacency data. Alternatively, we can approximate the integral using the Borel-Pad\'e method. While this method is extremely powerful, it does involve the numerical evaluation of a one-dimensional Laplace integral (for a detailed exposition see \cite{Aniceto:2018bis}). Remarkably, at no point do we need to compute the steepest descent and ascent manifolds in $\mathbb{C}^d$, which quickly becomes intractable as the dimension $d$ grows.

    \subsection{From adjacency to relevance: the South-East rule}
    The transseries \cref{eq:transseries} sums over all saddle points, weighted by the intersection numbers $n_j$. In practice, many complex saddle points are irrelevant ($n_j=0$), and including them leads to wrong, often exponentially growing, contributions. Determining the relevant saddle points is thus the first task. Traditionally, this requires tracing the steepest ascent manifolds $\mathcal{K}_j$ from each saddle point and counting their intersections with $\mathbb{R}^d$.

    The South-East rule replaces this geometric computation by a simple procedure in the Borel plane \cite{Aniceto:2026SouthEast}. Consider an exponent $f$ that is bounded from below on $\mathbb{R}^d$, as is the case for most oscillatory integrals in physics, and introduce a phase in the parameter $k \mapsto |k| e^{\ii\theta}$. At $\theta = \pi/2$, the integral becomes the non-oscillatory, Euclidean integral
    \begin{align}
        \Psi_{\theta=\pi/2} = \left(\frac{k}{2\pi}\right)^{d/2}\int_{\mathbb{R}^d} e^{-k f(\bm{x})}\dd\bm{x}\,.
    \end{align}
    The real plane $\mathbb{R}^d$ is now itself a union of steepest descent manifolds of the real critical points, so that the relevant saddle points are known exactly: the real critical points are relevant, with $n_j = 1$ in the normalization of \cref{eq:normalization}, and the complex critical points are irrelevant, with $n_j = 0$. We then rotate $\theta$ back from $\pi/2$ to the oscillatory integral at $\theta=0$. In the Borel plane, the singularities $t_j = -\ii\lambda_j$ rotate around the origin. Complex saddle points can only become relevant through Stokes transitions with saddle points that are already relevant. Such a transition occurs when two branch points line up horizontally in the Borel plane and the corresponding saddle points are adjacent, $K_{lj} \neq 0$ (compare with \cref{fig:StokesTransition}). At the transition, the relevant saddle point $\bm{x}_j$ changes the intersection number of the saddle point $\bm{x}_l$ by
    \begin{align}
        n_l \mapsto n_l + n_j K_{lj}\,.\label{eq:SouthEast}
    \end{align}
    Here, the signs of the Stokes constants and the intersection numbers refer to the orientation and normalization of the thimbles fixed in \cref{eq:normalization}, for which the real saddle points enter with $n_j = 1$.
    As the rotation is clockwise, the transitions are encountered in a definite order, and only the edges of the adjacency graph pointing from the North-West to the South-East (in the Borel plane at $\theta = 0$) can contribute. This yields the algorithm:
    \begin{enumerate}
        \item Select the edges $(t_j, t_l)$ of the adjacency graph pointing in the South-East direction, $\arg(t_l - t_j) \in [-\pi/2, 0]$, and order them from steep to shallow.
        \item Initialize $n_j = 1$ for the real critical points and $n_j = 0$ for the complex critical points.
        \item Loop over the ordered edges and apply the update \cref{eq:SouthEast} whenever $\bm{x}_j$ is relevant.
    \end{enumerate}
    The algorithm only requires the critical values and the Stokes constants, both of which follow from the Gauss-Manin system. Physically, the procedure resembles a Wick rotation: we start from the Euclidean integral, in which only the real saddle points contribute, and continue back to the oscillatory (Lorentzian) integral, keeping track of the complex saddle points that are switched on along the way. When $f$ is unbounded from below, the Euclidean integral diverges and complex saddle points may already be relevant at $\theta=\pi/2$. We refer to \cite{Aniceto:2026SouthEast} for the extension of the rule to this case.

    \subsection{Superasymptotics}
    Given the relevant saddle points, the transseries \cref{eq:transseries} is an asymptotic expansion of the moments $\bm{I}(k)$, the first component of which is the integral $\Psi$. The series around each saddle point diverge, as the coefficients $\bm{T}_m^{(j)}$ grow factorially. The resurgence relations make this growth precise: the late coefficients of the series of saddle point $\bm{x}_j$ are determined by the early coefficients of the saddle points adjacent to it. For non-degenerate saddle points, for which $\mu_j = 0$ in our normalization, they read
    \begin{align}
        \bm{T}_m^{(j)} \approx -\sum_{l} \frac{K_{lj}}{2\pi\ii}\sum_{r\geq 0} \frac{\Gamma(m - r)}{F_{jl}^{\,m-r}}\, \bm{T}_r^{(l)}\,, \quad m \gg 1\,,\label{eq:lateTerms}
    \end{align}
    with the singulant $F_{jl} = t_l - t_j$, the separation of the Borel singularities \cite{Berry:1991, Howls:1997}. The minus sign reflects our convention $\mathcal{S}^+ - \mathcal{S}^- = K_{lj}\bm{I}^{(l)}$ for the Stokes constants, and non-zero exponents $\mu_j$ shift the arguments of the Gamma functions. For the parabolic umbilic of \cref{sec:D5}, the relation \cref{eq:lateTerms} reproduces the coefficients $\bm{T}_m^{(j)}$ of all five saddle points to six significant digits for $m = 40, 60, 80$. Only the saddle points with $K_{lj} \neq 0$ appear in the sum. Consequently, it is not the nearest singularity in the Borel plane that controls the divergence of the series, but the nearest \textit{adjacent} singularity: a non-adjacent saddle point is invisible to the series, however close its critical value.

    By \cref{eq:lateTerms}, the terms $|\bm{T}_m^{(j)}|/|k|^m$ first decrease and then grow, reaching their smallest value near
    \begin{align}
        N_j \approx \alpha_j^{(0)} |k|\,, \quad \alpha_j^{(0)} = \min_{l:\, K_{lj}\neq 0} |F_{jl}|\,.
    \end{align}
    Truncating each series at its smallest term yields the superasymptotic approximation
    \begin{align}
        \bm{I}(k) \approx \sum_j n_j e^{\ii k\lambda_j} (\ii k)^{\mu_j}\sum_{m=0}^{N_j - 1}\frac{\bm{T}_m^{(j)}}{k^m}\,,
    \end{align}
    with an error that is exponentially small, of the order $e^{-|k F_{jl}|}$ relative to the leading contribution of saddle point $\bm{x}_j$ \cite{Berry:1991}. Near a Stokes line, where an exponentially subdominant contribution switches on, the jump in $n_l$ is not discontinuous but smoothed by an error function (see, for example, \cite{nemes2022dingle}), and the superasymptotic approximation remains accurate across the transition \cite{Berry:1991}. For large $k$, the superasymptotic approximation is often sufficient in practice. Its accuracy is, however, fundamentally limited by the exponentially small remainder.

    \subsection{Hyperasymptotics}\label{sec:hyperasymptotics}
    The remainder after optimal truncation is not lost. Berry and Howls showed that it can be written exactly as a sum over the adjacent saddle points, in which each saddle point $\bm{x}_l$ contributes its Stokes constant $K_{lj}$ times its own asymptotic series \cite{Berry:1991, Howls:1997}. Writing $S_j(k) = e^{-\ii k \lambda_j}\Psi^{(j)}(k) \sim \sum_m T_m^{(j)} k^{-m}$, the remainder after $N$ terms reads
    \begin{align}
        R_N^{(j)}(k) = -\sum_{l} \frac{K_{lj}}{2\pi\ii}\, (kF_{jl})^{-N} \int_0^\infty \frac{v^{N-1} e^{-v}}{1 - v/(kF_{jl})}\, S_l\!\left(\frac{v}{F_{jl}}\right) \dd v\,.\label{eq:exactRemainder}
    \end{align}
    Expanding the denominator in powers of $v/(kF_{jl})$ reproduces the late-term relation \cref{eq:lateTerms}. Substituting for $S_l$ its truncated series yields the first hyperasymptotic level,
    \begin{align}
        R_N^{(j)}(k) \approx -\sum_{l} \frac{K_{lj}}{2\pi\ii} \sum_{s=0}^{N_l - 1} T_s^{(l)} F_{jl}^{\,s}\, \frac{\Gamma(N-s)}{(kF_{jl})^{N}}\, h(N - s, kF_{jl})\,,\label{eq:levelOne}
    \end{align}
    in terms of the level-one hyperterminant
    \begin{align}
        h(M, w) = \frac{1}{\Gamma(M)}\int_0^\infty \frac{v^{M-1} e^{-v}}{1 - v/w}\,\dd v = -w\, e^{-w} E_M(-w)\,,
    \end{align}
    with $E_M$ the generalized exponential integral. This closed form is equivalent to the representation of the hyperterminant $F^{(1)}$ of Olde Daalhuis by the incomplete Gamma function \cite{Daalhuis:1998}. The function $h$ tends to one for large $|w|$ and has its branch cut on the positive real axis, which corresponds to the Stokes lines, where the median value applies. Substituting for $S_l$ its truncated series plus its own remainder \cref{eq:exactRemainder}, the approximation proceeds in levels. The level-$L$ term involves chains of adjacent saddle points $\bm{x}_j \to \bm{x}_{l_1} \to \dots \to \bm{x}_{l_L}$, weighted by the products of Stokes constants $K_{l_1 j}K_{l_2 l_1}\cdots$ and by multiple integrals, the hyperterminants of higher level \cite{Daalhuis:1996, Daalhuis:1998}. The hyperasymptotic expansion is thus organized by paths in the adjacency graph. Olde Daalhuis showed that the remainder at level $L$ is of the order $e^{-\alpha_j^{(L)}|k|}$, where $\alpha_j^{(L)}$ is the length of the shortest path of $L+1$ steps from $t_j$ in the adjacency graph, provided the series of $\bm{x}_j$ is truncated at $N_j = \alpha_j^{(L)}|k|$ and each re-expanded series at $N_{l} = \max(0, N_j - |kF_{jl}|)$ \cite{Daalhuis:1998b}. Each level thus improves on the previous one by an exponentially small factor, and the integral can be approximated to any desired precision, limited only by the computational effort.

    The higher-level hyperterminants admit the integral representation \cite{Daalhuis:1998}
    \begin{align}
        F^{(l+1)}\left(z; \begin{smallmatrix} M_0, \dots, M_l\\ \sigma_0, \dots, \sigma_l\end{smallmatrix}\right) &= \prod_{i=0}^l\Gamma(M_i) \int_0^\infty\!\!\cdots\!\int_0^\infty e^{-z s_0} (s_1 - s_0 - \sigma_0)^{-M_0}\cdots \nonumber\\
        &\qquad\times (s_l - s_{l-1} - \sigma_{l-1})^{-M_{l-1}} (-s_l - \sigma_l)^{-M_l}\,\dd s_l \cdots \dd s_0\,,
    \end{align}
    which is a Laplace transform in $s_0$ of a function with singularities at the cumulative singulants $-\sigma_0, -(\sigma_0 + \sigma_1), \dots$. For integer $M_i$, as in our normalization where the residues of the Fuchsian system are nilpotent, these Borel functions are hyperlogarithms with letters at the cumulative singulants. For instance, for the level-two hyperterminant with $M_0 = M_1 = 1$ we find
    \begin{align}
        F^{(2)}\left(z; \begin{smallmatrix} 1, 1\\ \sigma_0, \sigma_1\end{smallmatrix}\right) = \int_0^{\infty} e^{-zu}\, \frac{\log\big((u + \sigma_0)/\sigma_1\big) + \ii\pi}{u + \sigma_0 + \sigma_1}\,\dd u\,,
    \end{align}
    along suitably rotated contours, which we verified numerically. The integrand is the derivative of a weight-two hyperlogarithm with the letters $-\sigma_0$ and $-(\sigma_0 + \sigma_1)$. This connects the hyperterminants to the hyperlogarithmic moulds of \'Ecalle's resurgence monomials \cite{Ecalle:1981, Sauzin:2007mould}, and to the hyperlogarithm expansion of the Fuchsian system in \cref{sec:continuation}: the same iterated integrals that encode the Stokes constants also encode the hyperasymptotic remainders.

    The input required at every level is precisely the data computed in this paper: the critical values, the asymptotic series of all saddle points, and the Stokes constants. Whereas previous applications of hyperasymptotics to multidimensional integrals assumed the relevant saddle points and their adjacency to be known a priori \cite{Howls:1997, Delabaere:2002}, the Gauss-Manin system combined with the South-East rule determines both algorithmically. We have implemented the first two levels in \verb|IntegralResurgence.jl|, with the level-one hyperterminants in closed form and the second level by quadrature.

    \section{Examples}\label{sec:Example}
    We illustrate these methods with two examples. The parabolic umbilic diffraction integral is a genuinely two-dimensional integral with a polynomial exponent that is unbounded from below. The Lorentzian lens is a one-dimensional integral with a rational exponent, which we first lift to a polynomial representation in three dimensions.

    \subsection{The parabolic umbilic}\label{sec:D5}
    The parabolic umbilic ($D_5$) is the simplest catastrophe of codimension four that requires two state variables \cite{Poston:1996, Nye:1999}. We consider the canonical diffraction integral
    \begin{align}
        \Psi(x,y,z,w) = \frac{k}{2\pi\ii}\int_{\mathbb{R}^2} e^{\ii k f(u,v)}\dd u\,\dd v\,, \quad
        f(u,v) = u^4 + u v^2 + w u^2 + z v^2 + y v + x u\,,
    \end{align}
    with the state variables $(u,v)$ and the control parameters $(x,y,z,w)$. The term $uv^2$ renders the exponent unbounded from below, as $f\to-\infty$ when $v \to \pm\infty$ for $u < -z$. The critical points satisfy $v = -y/(2(u+z))$ and the quintic equation $(4u^3 + 2wu + x)\,4(u+z)^2 + y^2 = 0$, such that the exponent has five critical points, in agreement with the Milnor number $\mu = 5$ of the germ $u^4 + uv^2$. We select the generic configuration $(x,y,z,w) = (-\tfrac{5}{4}, \tfrac{1}{4}, 1, 0)$, for which the five critical points are non-degenerate with distinct critical values. Three critical points are real, a minimum $\bm{x}_1$ and two saddle points $\bm{x}_4$ and $\bm{x}_5$, and two form a complex conjugate pair,
    \begin{align}
        \begin{array}{lll}
            \bm{x}_1 \approx (0.678, -0.075)\,, & & t_1 \approx 0.646\,\ii\,,\\
            \bm{x}_{2,3} \approx (-0.342 \pm 0.586\,\ii,\, -0.106 \pm 0.094\,\ii)\,, & & t_{2,3} \approx \mp 0.539 - 0.305\,\ii\,,\\
            \bm{x}_4 \approx (-0.942, -2.142)\,, & & t_4 \approx -1.696\,\ii\,,\\
            \bm{x}_5 \approx (-1.052, 2.429)\,, & & t_5 \approx -2.840\,\ii\,,
        \end{array}
    \end{align}
    with the Borel singularities $t_j = -\ii f(\bm{x}_j)$.

    \paragraph{Gauss-Manin system.}
    The \verb|gmspoly.lib| library yields the good basis $\bm{\phi} = (1,\, 1+2u,\, v,\, -1+4u+8u^2,\, -7+4u+8v^2-16u^2)$ and the matrix
    \begin{align}
        B = \mathrm{diag}\left(\tfrac{5}{8}, \tfrac{7}{8}, 1, \tfrac{9}{8}, \tfrac{11}{8}\right).
    \end{align}
    These are the weights of the forms $\dd u\,\dd v$, $u\,\dd u\,\dd v$, $v\,\dd u\,\dd v$, $u^2\dd u\,\dd v$ and $v^2\dd u\,\dd v$ under the quasi-homogeneous rescaling of the germ $u^4 + uv^2$ with the weights $(w_u, w_v) = \left(\tfrac{1}{4}, \tfrac{3}{8}\right)$ (see \cref{ap:twisted}). As for the Airy integral, the unfolding terms only enter the matrix $M$, whose eigenvalues are the critical values $\lambda_j$.

\begin{figure}
        \centering
        \includegraphics[width=0.35\textwidth]{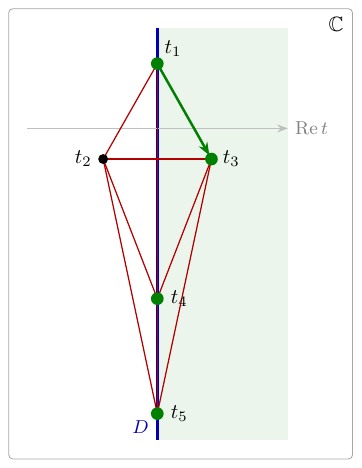}
        \caption{Adjacency graph of the parabolic umbilic for $(x,y,z,w) = (-\tfrac{5}{4}, \tfrac{1}{4}, 1, 0)$ in the Borel plane. The points are the Borel singularities $t_j = -\ii f(\bm{x}_j)$ of the five saddle points, and the red edges connect adjacent saddle points. All Stokes constants satisfy $|K_{lj}| \in \{0, 1\}$, and all pairs of saddle points are adjacent except $\bm{x}_4$ and $\bm{x}_5$. The edges from $t_1$ to $t_4$ and $t_5$ run along the imaginary axis. As the exponent is unbounded from below, the real plane maps to the whole imaginary axis $D$ (blue), and relevant complex saddle points lie in the shaded half-plane to its right. The South-East rule identifies the relevant saddle points (green): the three real saddle points and the complex saddle point $\bm{x}_3$, which is switched on by the Stokes transition along the edge from $t_1$ to $t_3$ (green arrow). Its complex conjugate $\bm{x}_2$ (black) is irrelevant.}\label{fig:ParabolicUmbilic}
    \end{figure}

    \paragraph{Adjacency.}
    Continuing the solutions of the Fuchsian system along the keyhole contours, we obtain the Stokes constants
    \begin{align}
        K =
        \begin{pmatrix}
             0 &  1 & -1 & -1 & -1\\
            -1 &  0 & -1 & -1 & -1\\
             1 &  1 &  0 & -1 & -1\\
             1 &  1 &  1 &  0 &  0\\
             1 &  1 &  1 &  0 &  0
        \end{pmatrix},
    \end{align}
    which are integers to a precision of $10^{-15}$. All pairs of saddle points are adjacent, except for the real saddle points $\bm{x}_4$ and $\bm{x}_5$ (see \cref{fig:ParabolicUmbilic}).

    \paragraph{Relevance.}
    As the exponent is unbounded from below, the real plane maps to the whole imaginary axis $D$ in the Borel plane, and the Euclidean integral at $\theta = \pi/2$ diverges. Following \cite{Aniceto:2026SouthEast}, we regulate the integral by adding $\epsilon v^4$ to the exponent, which renders it bounded from below. The regulator introduces four auxiliary critical points, which move to $t = +\ii\infty$ as $\epsilon \to 0$. As the original real saddle points $\bm{x}_4$ and $\bm{x}_5$ lie at $|v| \approx 2.3$, the regulator has to be small, $\epsilon \lesssim 1/128$, to leave the configuration of the original critical points intact. For $\epsilon = 1/128$ and $1/256$, the two real auxiliary critical points lie on the imaginary axis at $|t| \approx 45$ and $390$, respectively, and the two complex auxiliary critical points lie even further out. Evaluating the Stokes constants of the regulated exponent, we find that the real auxiliary critical points are only adjacent to the real saddle points $\bm{x}_4$ and $\bm{x}_5$. The corresponding edges are vertical and correspond to Stokes transitions at $\theta = \pi/2$ itself, which are accounted for by the initialization of the South-East rule. The complex auxiliary critical points are never switched on. Consequently, the regulator does not affect the relevance of the original critical points, and we may apply the South-East rule directly to the adjacency graph of \cref{fig:ParabolicUmbilic}.

    We initialize $n_1 = n_4 = n_5 = 1$ for the real saddle points and $n_2 = n_3 = 0$ for the complex saddle points. The complex saddle point $\bm{x}_2$ lies to the left of $D$ and remains irrelevant. The only edge pointing in the South-East direction from a relevant saddle point is the edge from $t_1$ to $t_3$, at the angle $\arg(t_3 - t_1) \approx -60.4^\circ$, which switches on the complex saddle point, $n_3 = n_1 K_{31}
         = 1$.
    The edges from $t_1$ to $t_4$ and $t_5$ are vertical, and the remaining South-East edges, $t_2 \to t_3$, $t_2 \to t_4$ and $t_2 \to t_5$, start at the irrelevant saddle point $\bm{x}_2$. We conclude that
    \begin{align}
        \Psi = \Psi^{(1)} + \Psi^{(3)} + \Psi^{(4)} + \Psi^{(5)}\,,
    \end{align}
    with the intersection numbers $\bm{n} = (1, 0, 1, 1, 1)$. The complex saddle point $\bm{x}_3$ contributes an oscillating term that is exponentially suppressed by $|e^{\ii k\lambda_3}| = e^{-k\,\mathrm{Im}\,\lambda_3} \approx e^{-0.54k}$.

    We verified this result numerically. The integral over $v$ is Gaussian and can be evaluated exactly, reducing the parabolic umbilic integral to the one-dimensional integral
    \begin{align}
        \Psi = \frac{k}{2\pi\ii}\int \sqrt{\frac{\pi}{-\ii k(u+z)}}\, e^{\ii k g(u)}\dd u\,, \quad g(u) = u^4 + wu^2 + xu - \frac{y^2}{4(u+z)}\,,
    \end{align}
    whose critical points and critical values coincide with those of $f$. The exponent $g$ has an essential singularity at $u = -z$, which we avoid by indenting the contour into the upper half-plane, where the integrand decays. A least-squares fit of the exact integral to the thimble integrals $\Psi^{(j)}$ over $k\in[3,8]$ yields $\bm{n} = (1,0,1,1,1)$ with a relative residual of $2\times 10^{-13}$.

\begin{figure}
        \centering
        \includegraphics[width=0.8\textwidth]{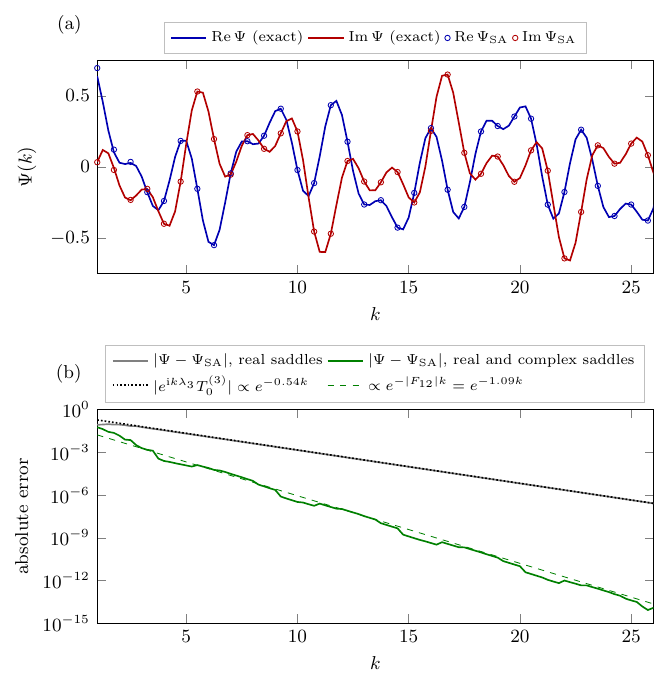}
        \caption{Superasymptotics of the parabolic umbilic integral for $(x,y,z,w) = (-\tfrac{5}{4}, \tfrac{1}{4}, 1, 0)$. (a) The real (blue) and imaginary (red) parts of the exact integral $\Psi(k)$ (lines) and of the superasymptotic approximation $\Psi_{\mathrm{SA}}(k)$ (circles), which includes the relevant saddle points $\bm{x}_1$, $\bm{x}_3$, $\bm{x}_4$ and $\bm{x}_5$, each truncated at $N_j = \lceil |F_j| k\rceil$. (b) The absolute error of the superasymptotic approximation with (green) and without (grey) the complex saddle point $\bm{x}_3$. Without $\bm{x}_3$, the error saturates at the size of its leading contribution, $|e^{\ii k\lambda_3} T_0^{(3)}| \propto e^{-0.54k}$ (dotted). With $\bm{x}_3$, the error decays as $e^{-|F_{12}|k} = e^{-1.09k}$ (dashed), set by the truncation of the series of the real minimum $\bm{x}_1$, until it reaches machine precision near $k = 26$.}\label{fig:D5Superasymptotics}
    \end{figure}

    \paragraph{Superasymptotics.}
    The asymptotic series of the saddle points illustrate that the divergence is governed by the nearest \textit{adjacent} singularity rather than the nearest singularity. The real saddle points $\bm{x}_4$ and $\bm{x}_5$ are each other's nearest singularity in the Borel plane, at the distance $|t_4 - t_5| \approx 1.15$, but are not adjacent. Estimating the singulant from the late coefficients with the root test $(\Gamma(m)/|T_m^{(j)}|)^{1/m}$ at $m \approx 100$, we find
    \begin{align}
        \begin{array}{cccc}
            \text{saddle point} & \text{late terms} & \text{nearest adjacent } |F_j| & \text{nearest singularity}\\
            \bm{x}_1 & 1.12 & 1.09 & 1.09\\
            \bm{x}_3 & 1.11 & 1.08 & 1.08\\
            \bm{x}_4 & 1.55 & 1.49 & 1.15\\
            \bm{x}_5 & 2.70 & 2.59 & 1.15
        \end{array}
    \end{align}
    where the root test converges slowly from above. The series of $\bm{x}_4$ and $\bm{x}_5$ are controlled by the adjacent complex saddle points $\bm{x}_2$ and $\bm{x}_3$, and are blind to each other. Truncating the series of the relevant saddle points at $N_j = \lceil |F_j|\,k\rceil$, with the distance $|F_j|$ to the nearest adjacent singularity, we obtain the superasymptotic approximation shown in \cref{fig:D5Superasymptotics}. Its error decreases as $e^{-|F_1|k} = e^{-1.09k}$, set by the real minimum $\bm{x}_1$, and reaches the machine precision $10^{-14}$ near $k = 26$. When omitting the complex saddle point $\bm{x}_3$, the error saturates at the size $e^{-0.54k}$ of its leading contribution.

    \subsection{The Lorentzian lens}\label{sec:lens}
    We next consider the one-dimensional Lorentzian lens model, given by the Kirchhoff-Fresnel integral
    \begin{align}
        \Psi(y) = \sqrt{\frac{k}{2\pi \mathrm{i}}} \int e^{\mathrm{i}k \left(
        \frac{(x-y)^2}{2} + \frac{1}{1+x^2}
        \right)}\mathrm{d}x\,.
    \end{align}
    Although this is a one-dimensional integral, the exponent is a rational rather than a polynomial function. Consequently, we first find a polynomial representation. Writing the exponential as a rational function $P(x) / Q(x)$, with $P(x) = (x-y)^2(1+x^2) + 2$ and $Q(x)=2(1+x^2)$, we insert Dirac delta function and write it as a Fourier integral to obtain the three-dimensional representation
    \begin{align}
        \Psi (y)
        &= \sqrt{\frac{k}{2\pi \mathrm{i}}} \int e^{\mathrm{i}k P(x_1)x_2} \delta(x_2 - 1/Q(x_1))\mathrm{d}x_1 \mathrm{d}x_2 \nonumber\\
        &= \frac{k}{2\pi} \int Q(x_1) e^{\mathrm{i}k \left(
        P(x_1)  x_2 + x_3  (x_2  Q(x_1) - 1)
        \right)}  \mathrm{d}x_1 \mathrm{d}x_2 \mathrm{d}x_3\,.
    \end{align}
    The Lorentzian lens integral has five saddle points. Using the \verb|Singular.jl| package, we find a good basis for $y=1$,
    \begin{align}
        \bm{\phi}(\bm{x}) =(1, x_1, -1/2 + 2 x_2, 2 x_3 + 4 x_2 - 2 x_1, 4 x_1 x_2)\,.
    \end{align}
    Performing Brieskorn reduction, we obtain the matrices 
    \begin{align}
        M = \begin{pmatrix}
             1/2 & -1 &  1  &  -1/2 &   0\\
   1   &  0  & 0   &  1/2  &  1\\
  5/4  & 1 & 1/2 &  3/4 & -1/2\\
 -3/2 &  0  & 1  &    0 &    1/2\\
  -2   &  2 &  2  &    0  &    0
        \end{pmatrix}, \quad
        B = \begin{pmatrix}
            1  & 0  & 0  & 0  &  0\\
0  & 1  & 0  & 0  &  0\\
0  & 0  & 2  & 0  &  0\\
0  & 0  & 0  & 3/2 & 0\\
0  & 0  & 0  & 0  &  2
        \end{pmatrix}.
    \end{align}

\begin{figure}
        \centering
        \includegraphics[width=0.6 \textwidth]{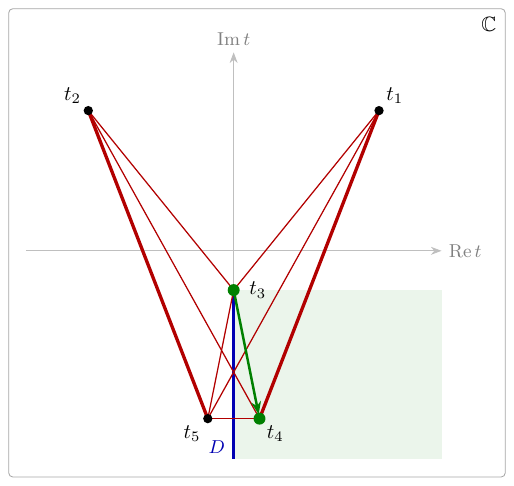}
        \caption{Adjacency graph of the Lorentzian lens for $y=1$ in the Borel plane. The points are the Borel singularities $t_j = -\ii f(x_j)$ of the five saddle points, and the red edges connect adjacent saddle points, with the line width proportional to $|K_{lj}| \in \{1, 2\}$. All pairs are adjacent except $\bm{x}_1$ and $\bm{x}_2$. As the exponent is bounded from below by its minimum $f(x_3)$, the real line maps to the half-line $D$ (blue) running downwards from $t_3$, and relevant complex saddle points lie in the shaded region to its right. The South-East rule identifies the relevant saddle points (green): the real saddle point $\bm{x}_3$ and the complex saddle point $\bm{x}_4$, which is switched on by the Stokes transition along the edge from $t_3$ to $t_4$ (green arrow).}\label{fig:Adjacency}
    \end{figure}

    Evaluating the Stokes constants with the Fuchsian system, we find the integer matrix
    \begin{align}
        K =
        \begin{pmatrix}
             0  & 0  &-1 &  2 & -1\\
             0  & 0  & 1 &  1 & -2\\
             1  &-1  & 0 & -1 &  1\\
            -2  &-1  & 1 &  0 &  1\\
             1  & 2  &-1 & -1 &  0
        \end{pmatrix},
    \end{align}
    with the entry $K_{lj}$ in row $l$ and column $j$, where the thimbles of the one-dimensional integral are normalized by $1/\sqrt{f''(x_j)}$, following the convention of \cref{sec:approximation}.
    We illustrate this graph in \cref{fig:Adjacency}. All pairs of saddle points are adjacent, except for the pair $\bm{x}_1$ and $\bm{x}_2$.

    \paragraph{Relevance.}
    We next apply the South-East rule of \cref{sec:approximation} to determine which saddle points contribute to the integral. The exponent is bounded from below on the real line, attaining its minimum $\mathcal{M} = f(\bm{x}_3) \approx 0.4155$ at the only real critical point $\bm{x}_3$, with $x_1 \approx 1.342$. In the Borel plane, the real line maps to the half-line $D$ running from $t_3 = -\ii\mathcal{M}$ to $-\ii\infty$ (see \cref{fig:Adjacency}). As the steepest ascent manifolds map to horizontal lines running to the left, relevant complex saddle points need to lie to the right of $D$. Of the four complex saddle points, only $\bm{x}_4$ satisfies this necessary condition.

    The adjacency graph contains five edges pointing in the South-East direction. Ordered from steep to shallow, these are
    \begin{align}
        t_3 \to t_4\ (-78.6^\circ)\,, \quad
        t_2 \to t_5\ (-68.8^\circ)\,, \quad
        t_2 \to t_4\ (-60.9^\circ)\,, \quad
        t_2 \to t_3\ (-51.0^\circ)\,, \quad
        t_5 \to t_4\ (0^\circ)\,,
    \end{align}
    with the angle $\arg(t_l - t_j)$ in brackets. We initialize the intersection numbers with $n_3 = 1$ for the real saddle point and $n_1=n_2=n_4=n_5=0$ for the complex saddle points, corresponding to the Euclidean integral at $\theta = \pi/2$. Rotating back to $\theta=0$, the first edge connects the relevant saddle point $\bm{x}_3$ to $\bm{x}_4$. By \cref{eq:SouthEast}, this Stokes transition switches on the complex saddle point, $n_4 = n_3 K_{43} = 1$. The remaining edges all start at the irrelevant saddle points $\bm{x}_2$ and $\bm{x}_5$ and leave the intersection numbers unchanged. We conclude that the Lorentzian lens integral decomposes as
    \begin{align}
        \Psi = \Psi^{(3)} + \Psi^{(4)}\,,
    \end{align}
    with the intersection numbers $\bm{n} = (0,0,1,1,0)$. The real saddle point $\bm{x}_3$ corresponds to the single real image of the lens. The complex saddle point $\bm{x}_4$ adds an oscillating contribution that is exponentially suppressed by the factor $|e^{\ii k\lambda_4}| = e^{-k\,\mathrm{Im}\,\lambda_4} \approx e^{-0.274 k}$, which can be interpreted as an evanescent complex image. The complex conjugate saddle point $\bm{x}_5$ would contribute an exponentially growing term and is, as expected, irrelevant, as are the saddle points $\bm{x}_1$ and $\bm{x}_2$.

    We verified this result numerically. In the original one-dimensional representation, the thimble integrals $\Psi^{(j)}$ can be evaluated directly by solving $f(x) = \lambda_j + \ii u^2$ for $x(u)$ and integrating over $u\in\mathbb{R}$, while the original integral is evaluated along a contour that coincides with the real line near the lens and is tilted by $45^\circ$ for large $|x|$. A least-squares fit of $\Psi = \sum_j n_j \Psi^{(j)}$ over $k \in [4,9]$ yields $\bm{n} = (0,0,1,1,0)$ with a relative residual of $10^{-13}$, where each thimble is oriented such that its leading term is $e^{\ii k\lambda_j}/\sqrt{f''(x_j)}$ with the principal branch of the square root.

    \paragraph{Superasymptotics.}
    The adjacency graph also determines the optimal truncation of the asymptotic series. The nearest saddle points adjacent to $\bm{x}_3$ are $\bm{x}_4$ and $\bm{x}_5$, both at the distance $|F_{34}| = |F_{35}| \approx 1.39$ in the Borel plane, so that the series of $\bm{x}_3$ is optimally truncated near $N_3 \approx 1.39 k$, with a relative error of the order $e^{-1.39k}$. The series of the complex saddle point $\bm{x}_4$ is limited by its adjacent neighbour $\bm{x}_5$ at the distance $|F_{45}| \approx 0.55$, giving $N_4 \approx 0.55k$ and an absolute error of the order $e^{-0.27k}e^{-0.55k} \approx e^{-0.82k}$, which dominates the error of the superasymptotic approximation. Note that for any fixed truncation order $m$, the algebraic error $\mathcal{O}(k^{-m})$ of the series of the real image eventually swamps the exponentially small contribution of the complex image, of the order $e^{-0.27k}$ (for three terms, beyond $k \approx 48$). Only the optimally truncated series, whose error decreases exponentially, resolves the complex image uniformly in $k$.

\begin{figure}
        \centering
        \includegraphics[width=0.8\textwidth]{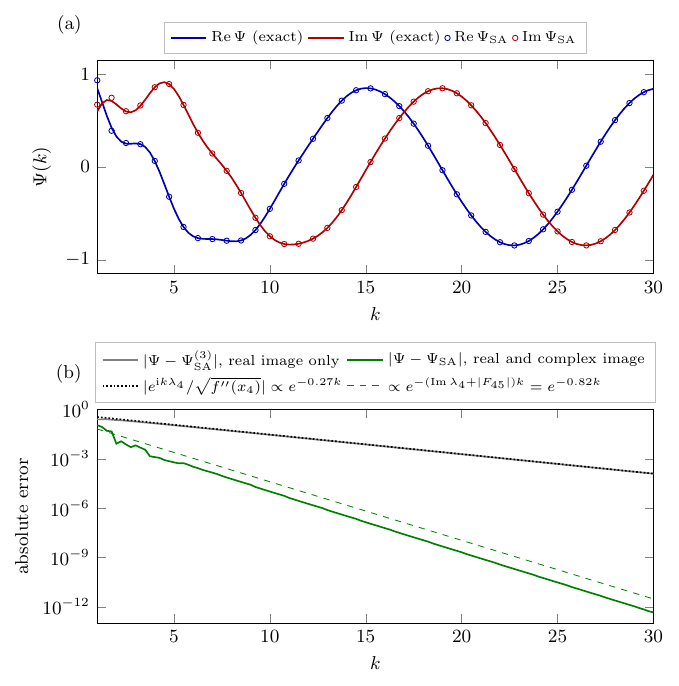}
        \caption{Superasymptotics of the Lorentzian lens integral for $y=1$. (a) The real (blue) and imaginary (red) parts of the exact integral $\Psi(k)$ (lines) and of the superasymptotic approximation $\Psi_{\mathrm{SA}}(k)$ (circles), which includes the real image $\bm{x}_3$ and the complex image $\bm{x}_4$, each truncated at $N_j = \lceil |F_j| k\rceil$. (b) The absolute error of the superasymptotic approximation with (green) and without (grey) the complex saddle point $\bm{x}_4$. Without $\bm{x}_4$, the error saturates at the size of its leading contribution, $|e^{\ii k\lambda_4}/\sqrt{f''(x_4)}| \propto e^{-0.27k}$ (dotted). With $\bm{x}_4$, the error follows the predicted decay $e^{-(\mathrm{Im}\,\lambda_4 + |F_{45}|)k} = e^{-0.82k}$ (dashed), set by the truncation of the series of $\bm{x}_4$, which is limited by the adjacent saddle point $\bm{x}_5$.}\label{fig:LensSuperasymptotics}
    \end{figure}

    We compute the coefficients of the asymptotic series of $\bm{x}_3$ and $\bm{x}_4$ to high order, and compare the superasymptotic approximation
    \begin{align}
        \Psi_{\mathrm{SA}}(k) = \sum_{j=3,4} e^{\ii k \lambda_j} \sum_{m=0}^{N_j-1} \frac{T_m^{(j)}}{k^m}\,, \quad N_j = \lceil |F_j|\, k \rceil\,,
    \end{align}
    with $|F_3| = |F_{34}|$ and $|F_4| = |F_{45}|$, to the exact integral in \cref{fig:LensSuperasymptotics}. The approximation is indistinguishable from the exact integral already for $k \gtrsim 2$. Its error decreases as $e^{-0.82k}$, in agreement with the estimate above, reaching $10^{-12}$ at $k=30$. When omitting the complex saddle point $\bm{x}_4$, the error saturates at the size $|e^{\ii k\lambda_4}/\sqrt{f''(x_4)}| \propto e^{-0.27k}$ of its leading contribution, confirming that the complex image identified by the South-East rule is genuinely part of the integral. The late coefficients confirm the role of adjacency: the coefficients of $\bm{x}_4$ grow as $T_m^{(4)} \sim \Gamma(m)/|F_{45}|^m$, controlled by the adjacent saddle point $\bm{x}_5$. Hyperasymptotics further reduces the error by resumming the remainders along the edges of the adjacency graph, starting with the edges $t_3 \to t_4$, $t_3\to t_5$ and $t_4 \to t_5$.

\section{A diagrammatic interpretation}
\label{sec:diagrams}

The mould expansion of the Fuchsian propagator admits a natural interpretation in terms of Feynman-like diagrams. Working in the eigenbasis of $M$, the propagator along a path $\gamma$ in the Borel plane takes the form
\begin{align}
    U_\gamma
    = I + \sum_{r=1}^{\infty} \sum_{j_1,\ldots,j_r}
    L^\gamma_{j_1\cdots j_r}\,
    \widetilde R_{j_r}\cdots\widetilde R_{j_1},
    \label{eq:diagrammatic_mould}
\end{align}
where the $L^\gamma_{j_1\cdots j_r}$ are iterated integrals of the logarithmic one-forms $\dd t/(t-t_j)$. In this basis, the residue matrices take the form $\widetilde R_j=E_{jj}\widetilde C$, with $\widetilde C=V^{-1}[B-(1+c)I]V$ and $E_{jj}$ the projector onto the $j$th saddle sector. Consequently, each nonzero matrix product describes an ordered sequence of transitions between saddle points.

\begin{figure}[t]
    \centering
    \begin{tikzpicture}[>=stealth, line width=0.9pt]
        \coordinate (a)  at (0,0);
        \coordinate (j1) at (2.5,0);
        \coordinate (j2) at (5,0);
        \coordinate (b)  at (7.5,0);

        \draw[->] (0.12,0) -- (2.38,0);
        \draw[->] (2.62,0) -- (4.88,0);
        \draw[->] (5.12,0) -- (7.38,0);

        \fill (a)  circle (2.5pt);
        \fill (j1) circle (2.5pt);
        \fill (j2) circle (2.5pt);
        \fill (b)  circle (2.5pt);

        \node[below=7pt] at (a)  {$a$};
        \node[below=7pt] at (j1) {$j_1$};
        \node[below=7pt] at (j2) {$j_2$};
        \node[below=7pt] at (b)  {$b$};

        \node[above=7pt] at (1.25,0) {$\widetilde C_{j_1a}$};
        \node[above=7pt] at (3.75,0) {$\widetilde C_{j_2j_1}$};
        \node[above=7pt] at (6.25,0) {$\widetilde C_{bj_2}$};
    \end{tikzpicture}
    \caption{A Feynman diagram for an ordered contribution to the Fuchsian propagator. The dots represent saddle sectors and the arrows carry elementary couplings from $\widetilde C$. The complete diagram is weighted by the corresponding hyperlogarithm, which encodes the ordering along the continuation path $\gamma$.}
    \label{fig:stokes_feynman}
\end{figure}
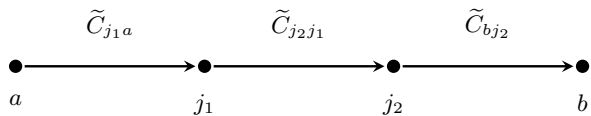

This suggests simple diagrammatic rules: the saddle points play the role of states, the matrix elements $\widetilde C_{ij}$ describe elementary couplings between them, and the hyperlogarithms $L^\gamma_{j_1\cdots j_r}$ supply the analytic weights of the ordered processes. For example, the diagram in \cref{fig:stokes_feynman} represents a contribution with weight
\begin{align}
    L^\gamma_{j_1j_2b}\,
    \widetilde C_{bj_2}\widetilde C_{j_2j_1}\widetilde C_{j_1a}.
\end{align}
The ordering matters: exchanging two intermediate saddle points generally changes the contribution, much as in time-ordered perturbation theory.

The Stokes constants inherit this diagrammatic interpretation. For continuation paths $\gamma_+$ and $\gamma_-$ passing on opposite sides of a Borel singularity, the discontinuity reads
\begin{align}
    U_{\gamma_+}-U_{\gamma_-}
    = \sum_{r=1}^{\infty}\sum_{j_1,\ldots,j_r}
    \operatorname{Disc} L_{j_1\cdots j_r}\,
    \widetilde R_{j_r}\cdots\widetilde R_{j_1}.
\end{align}
Projecting this discontinuity onto the local saddle solutions yields the Stokes constants. Their computation can therefore be viewed as a sum over ordered diagrams, weighted by discontinuities of hyperlogarithms.

For example, in the parabolic umbilic of \cref{sec:D5}, the saddle points $\bm{x}_4$ and $\bm{x}_5$ are not adjacent: $K_{45}=K_{54}=0$. Nevertheless, they can be connected by composite paths through other saddle sectors, such as $4\to1\to5$, $4\to2\to5$, and $4\to3\to5$. Such paths can contribute to the full transport operator even without a direct Stokes transition. The adjacency graph thus describes direct Stokes couplings, while the diagrammatic expansion reveals the richer structure of analytic continuation.

This provides a concrete realization of the picture anticipated by Berry and Howls \cite{Berry:1991}: saddle-point expansions behave as weakly interacting sectors, coupled through Stokes phenomena in the Borel plane.

    \section{Conclusion}\label{sec:Conclusion}
    In this paper, we developed an algebraic method to evaluate multidimensional oscillatory integrals through their asymptotic series. The resurgent approximation of an integral requires three pieces of global information: the asymptotic series of the saddle points, their adjacency, and their relevance. We showed that all three follow from integration by parts. The twisted de Rham cohomology of the exponent reduces the integral to a Gauss-Manin system of linear differential equations, whose size is set by the number of critical points rather than the dimension of the integral. The Gauss-Manin system yields the asymptotic series of all saddle points through a recursion. Its Borel transform is a Fuchsian system, whose analytic continuation yields the Stokes constants as integers, either by comparing two keyhole paths or with the single-loop formula. Combined with the critical values, the Stokes constants determine the relevant saddle points through the South-East rule, after which the series can be truncated optimally. Together, these steps take a polynomial exponent as input and return an exponentially accurate approximation of the integral, without computing a single steepest descent manifold.

    The examples illustrate the strengths of this approach. For the parabolic umbilic, a genuinely two-dimensional integral with an exponent unbounded from below, the method identified the adjacency of the five saddle points and the single relevant complex saddle point, and reproduced the integral to machine precision. The example also shows that the divergence of an asymptotic series is governed by the nearest \textit{adjacent} singularity: two real saddle points that are each other's nearest singularity in the Borel plane, but are not adjacent, are blind to each other. For the Lorentzian lens, the method identified the evanescent complex image that accompanies the real image, a contribution that is invisible in any fixed-order expansion at large $k$. We have made these methods available in the \verb|Julia| package \verb|IntegralResurgence.jl| (\url{https://github.com/jfeldbrugge/IntegralResurgence.jl}).

    The method has limitations that deserve further study. The Brieskorn reduction assumes a tame polynomial exponent; rational exponents require a polynomial lift, and exponents with critical points at infinity fall outside the present framework. The size of the Gauss-Manin system equals the number of critical points, which grows rapidly with the degree and the number of variables, and the computation of a good basis with \verb|Singular| becomes the bottleneck for large systems. For exponents that are unbounded from below, the South-East rule requires a regulator, whose auxiliary critical points must be tracked as the regulator is removed. Near caustics, where saddle points coalesce, the singulants become small and the superasymptotic approximation loses its accuracy, calling for uniform expansions in the Borel plane. Finally, we have only included the first two levels of hyperasymptotic approximation. As all of its ingredients, the series, the critical values and the Stokes constants, are provided by the Gauss-Manin system, an implementation with the hyperterminants of Olde Daalhuis \cite{Daalhuis:1996, Daalhuis:1998} is a natural next step and would extend the accuracy. From a mathematical perspective, the Stokes constants computed here are closely related to the intersection numbers of vanishing cycles in a distinguished basis, which suggests computing the monodromy groups and Coxeter-Dynkin diagrams of polynomials with these methods.

    \paragraph{Towards real-time path integrals.}
    The most exciting, and most speculative, extension concerns the real-time Feynman path integral, the infinite-dimensional limit of \cref{eq:OrigionalEquation} anticipated by Berry and Howls. In a path integral, the integration-by-parts identities underlying the Gauss-Manin system are the Schwinger-Dyson equations, which relate correlation functions, and the moments $\bm{I}^{(j)}$ play the role of a finite basis of correlators. In zero-dimensional quantum field theory, with a polynomial action in finitely many variables, our method applies without modification, and it yields the relevant classical solutions, their Stokes constants and the resummed perturbative series of every saddle point. For quantum mechanics, a time-sliced path integral with a polynomial potential is a finite-dimensional oscillatory integral of exactly the type studied here, with the classical paths as critical points and the classical actions as the singularities in the Borel plane. Two challenges stand out. First, the number of critical points of the discretized action grows exponentially with the number of time slices, so the Gauss-Manin system cannot be constructed naively; one would rather exploit the time-translation structure of the action, for instance by composing the Gauss-Manin systems of short time steps, in the spirit of the transfer matrix. Second, the continuum limit must be controlled: the critical values converge to the actions of the continuum classical solutions, but the Stokes constants and intersection numbers need not be stable under refinement of the discretization. If these challenges can be met, the adjacency of classical solutions would become a computable object, and the South-East rule would provide a precise version of the Wick rotation. For actions that are bounded from below, it rotates the Euclidean path integral, where only the real classical solutions contribute, back to the Lorentzian path integral, while tracking the complex classical solutions, such as instantons, that are switched on along the way \cite{Aniceto:2026SouthEast}. For actions that are unbounded from below, such as the Einstein-Hilbert action with its conformal factor problem \cite{Gibbons:1978, Hawking:1979, Mazur:1990}, the regulated South-East rule offers an algebraic criterion for which complex saddle points contribute, a question at the heart of the debate on the no-boundary proposal in Lorentzian quantum cosmology \cite{Feldbrugge:2017}. Even short of the continuum limit, the identification of relevant saddle points and their adjacency could guide the decomposition of path integrals onto Lefschetz thimbles in lattice simulations \cite{Cristoforetti:2013, Fujii:2013}, and real-time computations in quantum mechanics \cite{Feldbrugge:2023, Feldbrugge:2025PhRvD.111h3524F}.

    More than three decades after Berry and Howls envisioned hyperasymptotics for multiple integrals as a description of the ``feeble interaction of particles, represented by saddles'', the Gauss-Manin system makes this interaction computable: the Stokes constants are the couplings, the adjacency graph is the interaction network, and the South-East rule determines which particles are present. We hope that the methods and software presented here bring this vision closer to the oscillatory integrals and path integrals encountered in physics.

    \acknowledgments
    SC and JF thank Ines Aniceto, Chris Howls and Adri Olde Daalhuis for many illuminating discussions. SC thanks Gergo Nemes for many insightful discussions and is particularly grateful to Phil Trinh for originally introducing him to the fascinating problems of multidimensional steepest descent integrals five years ago; his encouragement helped inspire the present work. We also acknowledge the use of GPT-5.6 Sol and Claude Opus 4.8 for assistance with presentation and drafting, figure preparation, and numerical prototyping and cross-checks. All mathematical claims and results are the responsibility of the authors. J.F. is supported by a Postdoctoral Fellowship at National Taiwan University funded by the National Science and Technology Council (NSTC) 114-2923-M-002-011- MY5 Max Planck-IAS-NTU Center for Particle Physics, Cosmology and Geometry.

    \bibliographystyle{apsrev4-1}
    \bibliography{library.bib}

    \appendix
    \section{A linear system of ODEs with an irregular singularity at infinity}\label{ap:ODE}
    Assuming $M$ is diagonalizable, with the eigenvalue matrix $\Lambda = \text{diag}(\lambda_1, \dots, \lambda_n)$ and the orthonormal eigenvector matrix $V = (\bm{v}_1,\dots,\bm{v}_n)$, \textit{i.e.}, with $M \bm{v}_j = \lambda_j \bm{v}_j$ and $\bm{v}_i \cdot \bm{v}_j = \delta_{ij}$, we transform the Gauss-Manin system to
    \begin{align}
        \frac{\mathrm{d} \bm{Y} }{\mathrm{d}z } = \left[ \Lambda  
        +\frac{\tilde{B}}{z}
        \right] \bm{Y}\,,
    \end{align}
    with $\bm{I} = V \bm{Y}$, $z=\mathrm{i} k$ and the matrices $\tilde{B}=\frac{d}{2}I - V^{-1}BV$. This is known as a linear system of ODEs with an irregular singularity at infinity, as is manifest after a transformation $u = 1/z$. For large $z$, the first term on the right-hand side dominates over the second, leading to exponential growth/decay $\bm{Y} \propto e^{\Lambda z}$.
    For small $z$, the second term dominates over the first, given the power-law behaviour $\bm{Y} \propto  z^B$.
    The transseries combines the two into the ansatz,
    \begin{align}
        \bm{Y}(z) = e^{z \lambda} z^\mu \sum_{m=0}^\infty \frac{\bm{Y}_m}{z^m}\,,\label{eq:apAnsatz}
    \end{align}
    with the coefficients $\bm{Y}_m$ and scalars $\lambda$ and $\mu$. Substituting the ansatz in \cref{eq:apAnsatz}, we obtain the identity
    \begin{align}
        e^{z \lambda}z^\mu \left(
            \lambda  \sum_{m=0}^\infty \frac{\bm{Y}_m}{z^m}
            + 
            \sum_{m=0}^\infty \frac{(\mu-m) \bm{Y}_m}{z^{m+1}}
            \right)
        =
        e^{z \lambda}z^\mu \left(
            \Lambda \sum_{m=0}^\infty \frac{\bm{Y}_m}{z^m} 
            + 
            \sum_{m=0}^\infty  \frac{\tilde{B} \bm{Y}_m}{z^{m+1}}
            \right)\,.
    \end{align}
    Sorting in inverse powers of $z$ yields a couple realizations:
    \begin{itemize}
        \item To zeroth order in $z$,
        \begin{align}
            \Lambda \bm{Y}_0 = \lambda \bm{Y}_0\,,
        \end{align}
        indicating that $\lambda$ is an eigenvalue of $\Lambda$. For convenience, we select the $j$th eigenvalue $\lambda=\lambda_j$ and corresponding eigenvector $\bm{Y}_0=\bm{Y}_0^{(j)}=\bm{e}_j$ (defined by $[\bm{e}_j]_i=\delta_{ij}$). From here on, we label the coefficients $\lambda, \mu,$ and $\bm{Y}_m$ with the index $j$.
        \item To first order in $z$, we obtain the identity 
        \begin{align}
            (\Lambda - \lambda_j I)\bm{Y}_1^{(j)} = \left(\mu_j I - \tilde{B}\right)\bm{Y}_0^{(j)}\,.
        \end{align} 
        When considering the $j$th row, we find the coefficient $\mu_j = \tilde{B}_{jj}$.
        \item To $m$th order in $k$ (with $m \geq 1$), 
        \begin{align}
            (\Lambda  - \lambda_j I)  \bm{Y}_m^{(j)} = \left((\mu_j - m + 1) I -   \tilde{B} \right) \bm{Y}_{m-1}^{(j)}\,.
        \end{align}
        The $i$th row, when $i \neq j$, gives us the recursion relation
        \begin{align}
            Y_{i,m}^{(j)} = \frac{1}{\lambda_j - \lambda_i}\left( \sum_{l\neq i}\tilde{B}_{il} Y_{l,m-1}^{(j)} +( \mu_i - \mu_j + m - 1) Y_{i, m-1}^{(j)} \right) \,. \label{eq:ap1}
        \end{align}
        The $j$th row, yields the identity
        \begin{align}
            Y_{j, m}^{(j)} = -\frac{1}{m}\sum_{l \neq j} \tilde{B}_{jl}Y_{l,m}^{(j)} \,.\label{eq:ap2}
        \end{align}
    \end{itemize}

    Starting from the eigenvalue $\lambda_j$, the associated eigenvector $\bm{Y}_0^{(j)}= \bm{e}_j$ and the coefficient $\mu_j$, we use \cref{eq:ap1,eq:ap2} to compute $\bm{Y}_m^{(j)}$. Moving from $\bm{Y}_{m}^{(j)}$ to $\bm{Y}_{m+1}^{(j)}$, we first update the rows $i \neq j$ using \cref{eq:ap1}. Subsequently, we find the $j$th row with \cref{eq:ap2}. We rotate the coefficients $\bm{Y}_m^{(j)}$ back to the original problem with the eigenvector matrix, $\bm{T}_m^{(j)} = \mathrm{i}^{-m} V \bm{Y}_m^{(j)}$. Finally, we rescale the coefficients such that the first row $[\bm{T}^{(j)}_{0}]_0= 1/\sqrt{\det \mathcal{H}f(\bm{x}_j)}$.
    
    \section{Twisted de Rham cohomology and Brieskorn reduction}\label{ap:twisted}
    In this appendix, we give a more formal account of the integration-by-parts identities used in the derivation of the Gauss-Manin system. The underlying structure is a pairing between two finite-dimensional vector spaces: the space of integration contours along which $e^{\ii k f}$ decays (twisted homology) and the space of integrands modulo total derivatives (twisted cohomology). Throughout, we assume that $f\in \mathbb{C}[\bm{x}]$ is a Morse polynomial with $n$ non-degenerate critical points $\bm{x}_j$ and distinct critical values $\lambda_j = f(\bm{x}_j)$. We moreover assume that $f$ is \textit{tame}, \textit{i.e.}, that $f$ has no critical points at infinity, meaning that $|\nabla f(\bm{x})|$ is bounded away from zero for large $|\bm{x}|$ \cite{Sabbah:1999}. For more details, we refer to \cite{Pham1965FormulesDP, pham2011singularities, ArnoldVG:1988, Schwarz:2009, Witten:2010} and, for the theory of twisted cohomology and its intersection pairing, to \cite{Aomoto:2011, Cho:1995}.

    \subsection{Twisted homology: the integration contours}
    Consider the integral $\int_\Gamma g(\bm{x}) e^{\ii k f(\bm{x})}\dd \bm{x}$ along a $d$-dimensional real contour $\Gamma \subset \mathbb{C}^d$. The integral converges when $\Gamma$ either is compact or extends to infinity only in regions where the integrand decays, \textit{i.e.}, where $\mathrm{Re}[\ii k f(\bm{x})] \to -\infty$. Two such contours give the same integral when they can be continuously deformed into one another while keeping their ends in the region of decay, as follows from the Cauchy theorem in several variables. Formally, the contours are classified by the relative homology group
    \begin{align}
        H_d(\mathbb{C}^d, \mathcal{B}_T)\,, \quad \text{with} \quad \mathcal{B}_T = \{ \bm{x}\in \mathbb{C}^d \,|\, \mathrm{Re}[\ii k f(\bm{x})] < -T\}\,,
    \end{align}
    for sufficiently large $T$, consisting of $d$-dimensional chains with boundary in $\mathcal{B}_T$, modulo deformations. For a tame Morse polynomial, this group is a lattice of rank $n$, generated by the Lefschetz thimbles $\mathcal{J}_1,\dots,\mathcal{J}_n$ \cite{Pham1965FormulesDP}. The thimble $\mathcal{J}_j$ is the union of the paths of steepest descent of the Morse function $h(\bm{x}) = \mathrm{Re}[\ii k f(\bm{x})]$ emanating from the saddle point $\bm{x}_j$. Along these paths, the phase $\mathrm{Im}[\ii k f(\bm{x})]$ is constant, so that the integrand does not oscillate.

    The ascent manifolds $\mathcal{K}_j$, consisting of the paths of steepest ascent from $\bm{x}_j$, generate the dual group with the region of decay replaced by the region of growth, $\mathrm{Re}[\ii k f(\bm{x})] > T$. For generic $k$, the two bases are dual with respect to the intersection number, $\langle \mathcal{J}_i, \mathcal{K}_j \rangle = \delta_{ij}$. As a consequence, any contour $\Gamma$ decomposes as
    \begin{align}
        \Gamma = \sum_{j=1}^n \langle \Gamma, \mathcal{K}_j \rangle\, \mathcal{J}_j\,.
    \end{align}
    For the original integration domain $\Gamma = \mathbb{R}^d$, this is the Picard-Lefschetz formula \cref{eq:PicardLefscehtz} with the intersection numbers $n_j = \langle \mathbb{R}^d, \mathcal{K}_j\rangle$.

    The thimbles depend on the phase of $k$. When $k = |k|e^{\ii\theta}$ is rotated, a thimble $\mathcal{J}_j$ changes discontinuously when a path of steepest descent from $\bm{x}_j$ runs into another saddle point $\bm{x}_l$. As the phase $\mathrm{Im}[\ii k f]$ is constant along the path, this requires $\ii k (\lambda_l - \lambda_j) \in \mathbb{R}_{<0}$. In terms of the Borel singularities $t_j = -\ii \lambda_j$, this is precisely the condition that $t_l$ lies on the Laplace ray $(t_j, t_j + e^{-\ii\theta}\infty)$ (see \cref{fig:StokesTransition}). Across this Stokes ray, the thimble jumps as $\mathcal{J}_j \mapsto \mathcal{J}_j + K_{lj} \mathcal{J}_l$, where, by the Picard-Lefschetz formula, $K_{lj}$ is (up to sign) the intersection number of $\mathcal{J}_j$ with $\mathcal{K}_l$ on the Stokes ray \cite{Pham1965FormulesDP, ArnoldVG:1988}. The Stokes constants computed in this paper are thus topological invariants: integers that count how the descent manifold of one saddle point wraps around the descent manifold of another.

    \subsection{Twisted cohomology: the integrands}
    We next consider the integrands. It is convenient to work with differential forms, writing the integrand as $e^{\ii k f}\omega$ with the algebraic $d$-form $\omega = g(\bm{x})\,\dd x_1\wedge\dots\wedge \dd x_d$. Differentiating the product with an algebraic $(d-1)$-form $\eta$, we obtain
    \begin{align}
        \dd\left(e^{\ii k f}\eta\right) = e^{\ii k f}\, \nabla_k \eta\,, \quad \text{with} \quad \nabla_k = \dd + \ii k\, \dd f \wedge\,.
    \end{align}
    The operator $\nabla_k$ is the exterior derivative \textit{twisted} by the exponential. It satisfies $\nabla_k^2 = 0$ and defines the twisted de Rham complex
    \begin{align}
        0 \to \Omega^0 \xrightarrow{\nabla_k} \Omega^1 \xrightarrow{\nabla_k} \cdots \xrightarrow{\nabla_k} \Omega^d \to 0\,,
    \end{align}
    with $\Omega^p$ the algebraic $p$-forms on $\mathbb{C}^d$, \textit{i.e.}, $p$-forms with polynomial coefficients. For the $(d-1)$-form
    \begin{align}
        \eta = \sum_{i=1}^d (-1)^{i-1} h_i(\bm{x})\, \dd x_1\wedge \dots \wedge \widehat{\dd x_i} \wedge \dots \wedge \dd x_d\,,
    \end{align}
    where the hat denotes omission, the twisted derivative reads
    \begin{align}
        \nabla_k \eta = \left[\nabla \cdot \bm{h}(\bm{x}) + \ii k\, \bm{h}(\bm{x}) \cdot \nabla f(\bm{x})\right]\dd x_1\wedge\dots\wedge \dd x_d\,.
    \end{align}
    These are exactly the total derivatives that integrate to zero in the main text. The space of integrands modulo total derivatives is the top twisted de Rham cohomology group
    \begin{align}
        H^d = \Omega^d / \nabla_k \Omega^{d-1}\,.
    \end{align}
    For a tame polynomial, the lower cohomology groups vanish and $H^d$ is $n$-dimensional, with $n$ the number of critical points, matching the rank of the twisted homology \cite{Sabbah:1999}.

    The integral defines a pairing between twisted homology and cohomology,
    \begin{align}
        \langle \Gamma, [\omega] \rangle = \int_\Gamma e^{\ii k f}\omega\,.
    \end{align}
    The pairing is well defined: shifting $\omega$ by a total derivative does not change the integral, as by Stokes' theorem
    \begin{align}
        \int_\Gamma e^{\ii k f}\nabla_k \eta = \int_{\partial \Gamma} e^{\ii k f}\eta = 0\,,
    \end{align}
    since the boundary $\partial\Gamma$ lies at infinity in the region where $e^{\ii k f}$ decays faster than any polynomial. Moreover, the pairing is non-degenerate, so that the $n\times n$ period matrix $P_{ij}(k) = \langle \mathcal{J}_j, [\phi_i \,\dd\bm{x}]\rangle$ is invertible for any basis $\bm{\phi}$ of $H^d$. This is the precise sense in which the $n$ standard moments $\bm{I}^{(j)}$ capture the full information about the integrals over the thimble $\mathcal{J}_j$: the moment of any polynomial $g$ is a linear combination of the standard moments with coefficients that do not depend on the contour.

    For physicists, this structure is familiar from the reduction of Feynman integrals to master integrals with integration-by-parts identities \cite{Chetyrkin:1981qh, Laporta:2000dsw}. In recent years, this reduction has been formulated in terms of twisted de Rham cohomology, where the decomposition onto master integrals follows from intersection numbers of twisted cocycles \cite{Mizera:2017rqa, Mastrolia:2018uzb, Frellesvig:2019kgj, Frellesvig:2019uqt, Frellesvig:2020qot, Brunello:2023rpq}; see \cite{Mizera:2020wdt, Weinzierl:2022eaz} for reviews. The twisted cohomology classes $[\phi_i\,\dd \bm{x}]$ play the role of the master integrals and the dimension $n$ counts the number of independent master integrals.

    \subsection{The Jacobi algebra and the large-$k$ limit}
    To make the dependence on $k$ explicit, we introduce the small parameter $\tau = 1/(\ii k)$ and write the identities as
    \begin{align}
        \bm{h}(\bm{x}) \cdot \nabla f(\bm{x}) \equiv -\tau\, \nabla \cdot \bm{h}(\bm{x})\,.\label{eq:apIBP}
    \end{align}
    In the limit $k\to\infty$, the right-hand side vanishes, and the identities reduce to $\bm{h}\cdot\nabla f \equiv 0$. Polynomials modulo this relation form the Jacobi algebra
    \begin{align}
        \mathrm{Jac}(f) = \mathbb{C}[\bm{x}] / (\partial_1 f, \dots, \partial_d f)\,.
    \end{align}
    This reflects the method of stationary phase: to leading order in $1/k$, the integral localises on the critical points, so that two polynomials that agree at all critical points yield the same leading-order contribution. For a Morse polynomial, the Jacobi algebra is $n$-dimensional and evaluation at the critical points identifies it with $\mathbb{C}^n$,
    \begin{align}
        g \mapsto \left(g(\bm{x}_1),\dots,g(\bm{x}_n)\right)\,.
    \end{align}
    In this representation, the multiplication by $f$ acts diagonally with the critical values $\lambda_j = f(\bm{x}_j)$ on the diagonal. This explains Stickelberger's theorem used in the main text: in any basis $\bm{\phi}$, the matrix $M$ of multiplication by $f$ in $\mathrm{Jac}(f)$ has the eigenvalues $\lambda_j$, with eigenvectors proportional to $\bm{\phi}(\bm{x}_j)$.

    The Jacobi algebra is the $\tau\to 0$ limit of the twisted cohomology. More precisely, let $G_0$ denote the space of $d$-forms with coefficients in $\mathbb{C}[\bm{x}][\tau]$ modulo the relations \cref{eq:apIBP}. This is the (Fourier transformed) Brieskorn lattice of $f$ \cite{Brieskorn:1970, Sabbah:1999}. For tame $f$, it is a free $\mathbb{C}[\tau]$-module of rank $n$, and setting $\tau=0$ gives
    \begin{align}
        G_0 / \tau G_0 \cong \mathrm{Jac}(f)\,\dd\bm{x}\,.
    \end{align}
    As a consequence, any basis $\bm{\phi}$ of the Jacobi algebra lifts to a basis of the twisted cohomology. This is why the basis functions in the main text can be chosen as monomials spanning the Jacobi algebra $\mathrm{Jac}(f)$.

    \subsection{Brieskorn reduction}
    Brieskorn reduction expresses an arbitrary polynomial $g$ in the basis $\bm{\phi}$. In the Jacobi algebra, we divide $g$ by the gradient of $f$,
    \begin{align}
        g(\bm{x}) = \bm{m}\cdot \bm{\phi}(\bm{x}) + \bm{\beta}(\bm{x}) \cdot \nabla f(\bm{x})\,,
    \end{align}
    with $\bm{m}\in \mathbb{C}^n$ and $\bm{\beta}\in \mathbb{C}[\bm{x}]^d$. In practice, the remainder $\bm{m}\cdot\bm{\phi}$ and the quotient $\bm{\beta}$ are obtained with polynomial division with respect to a Gr\"obner basis of the Jacobi ideal $(\partial_1 f,\dots,\partial_d f)$, for which $\bm{\phi}$ is the standard monomial basis. Applying \cref{eq:apIBP} to the second term gives
    \begin{align}
        g \equiv \bm{m}\cdot\bm{\phi} - \tau\, \nabla\cdot\bm{\beta}\,.
    \end{align}
    The divergence $\nabla \cdot \bm{\beta}$ is again a polynomial, which we reduce in the same way. Iterating, we obtain the normal form
    \begin{align}
        g \equiv \sum_{r=0}^\infty \tau^r\, \bm{m}_r \cdot \bm{\phi}\,,
    \end{align}
    which is the expansion of $\bm{\alpha}(k)$ in the main text. As the divergence lowers the degree of the polynomial, and $G_0$ is a free $\mathbb{C}[\tau]$-module for tame $f$, the normal form is a polynomial in $\tau$, \textit{i.e.}, the reduction terminates after finitely many steps. The number of steps depends on the polynomial $g$ and on the choice of basis $\bm{\phi}$.

    \subsection{The Gauss-Manin connection}
    The moments $J_i = \int_\Gamma \phi_i e^{\ii k f}\dd\bm{x}$ along any rapid-decay contour $\Gamma$ satisfy
    \begin{align}
        \frac{\dd J_i}{\dd k} = \ii \int_\Gamma f \phi_i\, e^{\ii k f}\dd\bm{x}\,.
    \end{align}
    The derivative with respect to $k$ thus acts on cohomology as multiplication by $\ii f$. Reducing the products $f\phi_i$ to normal form,
    \begin{align}
        f \phi_i \equiv \sum_{j=1}^n \left(M_{ij} - \sum_{r=1}^\infty \tau^r [B_r]_{ij}\right)\phi_j\,,
    \end{align}
    we obtain a linear system of ordinary differential equations for $\bm{J}$, known as the Gauss-Manin connection. The normalization $(k/2\pi\ii)^{d/2}$ of the moments in the main text adds the term $\frac{d}{2k}I$. The Gauss-Manin system has a regular singularity at $k=0$ and an irregular singularity of rank one at $k=\infty$, in terms of $\tau$ a pole of order two at $\tau=0$. The irregular singularity is responsible for the exponential factors $e^{\ii k\lambda_j}$ and the divergent asymptotic series studied in this paper.

    The matrices $B_r$ depend on the choice of basis. A change of basis $\bm{\phi} \mapsto P(\tau)\bm{\phi}$, with $P(\tau)$ an invertible matrix polynomial in $\tau$, acts on the connection as a gauge transformation. Saito showed that, for tame polynomials, there exists a \textit{good basis} in which the series truncates after the first term \cite{Saito:1989, Saito:2013, Sabbah:1999},
    \begin{align}
        f\phi_i \equiv \sum_{j=1}^n \left(M_{ij} - \tau B_{ij}\right)\phi_j\,,
    \end{align}
    with constant matrices $M$ and $B$. This yields the Gauss-Manin system \cref{eq:GaussManin}. The existence of such a basis is the solution of the so-called Birkhoff problem for the connection. In a good basis, the eigenvalues of $B$ are the spectral numbers of $f$, which generalize the scaling dimensions of the basis forms $\phi_i\,\dd\bm{x}$ \cite{Sabbah:1999}. A good basis is computed by the algorithm of Schulze implemented in the \verb|gmspoly.lib| library of \verb|Singular| \cite{Schulze:2003}.

    \paragraph{Quasi-homogeneous polynomials.}
    The matrix $B$ is easily computed for a quasi-homogeneous polynomial, \textit{i.e.}, when there exist weights $w_1,\dots,w_d > 0$ such that $f(s^{w_1}x_1,\dots,s^{w_d}x_d) = s f(\bm{x})$. Differentiating with respect to $s$ at $s=1$ gives the Euler relation $f = \sum_i w_i x_i\partial_i f$. For the monomial $\phi = \bm{x}^{\bm{a}} = x_1^{a_1}\cdots x_d^{a_d}$, we set $h_i = w_i x_i \phi$ in \cref{eq:apIBP} to obtain
    \begin{align}
        f\phi = \sum_{i=1}^d w_i x_i\phi\, \partial_i f
        \equiv -\tau\sum_{i=1}^d w_i \partial_i\left(x_i \bm{x}^{\bm{a}}\right)
        = -\tau\, \ell(\bm{a})\, \phi\,, \quad \text{with} \quad \ell(\bm{a}) = \sum_{i=1}^d w_i(a_i+1)\,.
    \end{align}
    The reduction terminates after a single step, so that any monomial basis is a good basis with $M=0$ and the diagonal matrix $B_{ii} = \ell(\bm{a}_i)$. The number $\ell(\bm{a})$ is the weight of the form $\bm{x}^{\bm{a}}\dd\bm{x}$ under the rescaling. For the Airy integral with $f = u^3$, the weight is $w = 1/3$ and the basis $\bm{\phi} = (1,u)$ gives $B = \mathrm{diag}(1/3, 2/3)$, in agreement with the main text. Turning on the external parameter, $f=u^3+xu$, breaks the quasi-homogeneity and generates the matrix $M$, while $B$ remains unchanged.

\end{document}